\documentclass[12pt,preprint]{aastex}
\slugcomment{Submitted to ApJ}
\shorttitle{Mass Outflows in RY Scuti}
\shortauthors{Grundstrom et al.}
\begin{document}


\title{A Spectroscopic Study of Mass Outflows in the Interacting Binary RY Scuti}

\author{Erika D. Grundstrom, Douglas R. Gies\altaffilmark{1}, 
Todd C. Hillwig\altaffilmark{1,2}, and M. Virginia McSwain\altaffilmark{1,3,4}}
\affil{Center for High Angular Resolution Astronomy and \\
 Department of Physics and Astronomy, Georgia State University, Atlanta, GA;
erika@chara.gsu.edu, gies@chara.gsu.edu, todd.hillwig@valpo.edu, mcswain@astro.yale.edu}

\author{Nathan Smith}
\affil{Astronomy Department, University of California, Berkeley, CA;
nathans@astro.berkeley.edu}

\author{Robert D. Gehrz}
\affil{Astronomy Department, University of Minnesota, Minneapolis, MN;
gehrz@astro.umn.edu}

\author{Otmar Stahl}
\affil{Landessternwarte Heidelberg, Heidelberg, Germany;
o.stahl@lsw.uni-heidelberg.de}

\author {Andreas Kaufer}
\affil{European Southern Observatory, Santiago, Chile;
akaufer@eso.org}

\altaffiltext{1}{Visiting Astronomer, Kitt Peak National Observatory and 
Cerro Tololo Interamerican Observatory, National 
Optical Astronomy Observatory, operated by the Association of Universities for 
Research in Astronomy (AURA), Inc., under cooperative agreement with the 
National Science Foundation.}
\altaffiltext{2}{Current address: Department of Physics and Astronomy, 
Valparaiso University, Valparaiso, IN.}
\altaffiltext{3}{Current address: Department of Astronomy, Yale University, 
New Haven, CT.}
\altaffiltext{4}{NSF Astronomy and Astrophysics Postdoctoral Fellow.}


\begin{abstract}

The massive interacting binary RY~Scuti is an important representative of 
an active mass-transferring system that is changing before our eyes
and which may be an example of the formation of a Wolf-Rayet star
through tidal stripping. 
Utilizing new and previously published spectra, we present examples of how 
a number of illustrative absorption and emission features vary during 
the binary orbit. 
We identify spectral features associated with each component, 
calculate a new, double-lined spectroscopic binary orbit, and find
masses of $7.1 \pm 1.2~M_\odot$ for the bright supergiant and 
$30.0 \pm 2.1~M_\odot$ for the hidden massive companion. 
Through tomographic reconstruction of the component spectra from the composite
spectra, we confirm the O9.7~Ibpe spectral class of the bright supergiant and
discover a B0.5~I spectrum associated with the hidden massive companion; 
however, we suggest that the latter is actually the spectrum of the photosphere
of the accretion torus immediately surrounding the massive companion. 
We describe the complex nature of the mass loss flows from the system in the
context of recent hydrodynamical models for $\beta$ Lyr, leading us to 
conclude RY~Scuti has matter leaving 
the system in two ways: 1) a bipolar outflow from winds generated by the 
hidden massive companion, and 2) an outflow from the bright O9.7~Ibpe 
supergiant in the region near the L2 point to fill out a large, dense 
circumbinary disk.  This circumbinary disk (radius $\approx$~1~AU) may feed the
surrounding double-toroidal nebula (radius $\approx$~2000~AU).

\end{abstract}



\setcounter{footnote}{4}

\section{Introduction}

RY~Scuti (HD~169515) is a distant (1.8 $\pm$ 0.1 kpc; \citealt{smi02})  
and massive eclipsing binary system with an orbital period currently estimated 
at 11.12445 days \citep{kre04}.  
Analysis of the light curve indicates that at least one of the components 
fills its Roche lobe.  The present configuration may be semi-detached 
\citep{cowhut76}, 
overcontact \citep*{mil81, djur01}, or one with the more massive component 
embedded in an opaque thick disk \citep{kin79,ant99}.
The binary appears to be in an advanced stage of evolution, and it has ejected 
gas into a young ($\sim$130 year old; \citealt*{smi01}), double-toroidal 
emission nebula.
The $\approx$ 2000 AU nebula is angularly resolved in radio images 
\citep{geh95}, infrared images \citep{geh01}, and in {\it Hubble Space 
Telescope} H$\alpha$ images \citep{smi99,smi01,smi02}.  Since the nebula is so
young and the star system is still in an active Roche-lobe overflow phase,
RY Scuti is a powerful laboratory for studying non-spherical mass and
angular momentum loss in interacting binaries.

Only one of the component stars is readily visible in optical spectra 
and it has a spectral classification of O9.7~Ibpe~var \citep{wal82}. 
We will refer to this component as the ``supergiant.''   The other star 
appears to be enshrouded since it is very difficult to detect in the spectrum.
Most investigators agree that this second star is the more massive of the two, 
and we will refer to the second component as the ``massive companion.''
However, the actual masses of the stars in this system are debatable.  
The radial velocity curve of the supergiant is reasonably well established, 
but the results for the massive companion depend critically on what
spectral features one assumes are associated with that star.  As these features
are difficult to observe and interpret, the estimated mass range has been huge.
For example, 
\citet{pop43} arrived at a total system mass in excess of $100 M_\odot$. 
Later investigators found lower values: 
\citet{cowhut76} estimated masses of 36 and $46 M_\odot$,
\citet{skul92} found 8 and $26 M_\odot$, 
and \citet*{sah02} estimated 9 and $36 M_\odot$
(for the supergiant and massive companion, respectively, in each case).
There are a number of important photometric studies (e.g., ranging from the 
discovery by \citealt{gap37} through photoelectric investigations by 
\citealt{giu81} and \citealt{mil81}, 
and up to the most recent multi-color work by \citealt{djur01}), however the 
results differ with regard to the assumed binary configuration and depend 
sensitively on the mass ratio adopted from spectroscopy. 

This unique system is representative of the short-lived,
active mass transfer stage in the evolution of massive binaries. 
Theoretical models \citep*{pet05} indicate most of the mass transfer 
occurs during a brief ($\approx 10^4$ yr), and thus rare, phase in which the 
mass donor transfers most of its mass to the mass gainer star.  Mass transfer 
will shrink the orbital 
dimensions until the mass ratio is reversed (and the gainer becomes the 
more massive star), and then the system will enter a slower (and longer lived) 
mass transfer phase as the binary expands.  The massive binaries that are just 
emerging from the rapid mass transfer phase probably belong to the observed 
class of W~Serpentis binaries \citep{tar00}.   Only the mass donor star is 
visible in 
the spectra of these binaries and the more massive gainer star is hidden in a 
thick accretion disk (one source of emission lines in the spectra).  The mass 
transfer process is complex and leaky, and a significant fraction of the mass 
loss leaves the system completely (as described by \citealt{har02} for the best
known object of the class, $\beta$~Lyr).  The mass donor may eventually lose 
its entire hydrogen envelope and emerge as a Wolf-Rayet star. Therefore a 
system like RY~Scuti may be the progenitor of a WR+O binary system 
\citep{giu81,ant88,smi02}.  

In a prior paper \citep{smi02}, several of us presented a detailed study
of the spectral features formed in the surrounding double-toroidal nebula through an examination 
of a set of high dispersion spectra obtained with the ESO FEROS spectrograph. 
Here we use the same set of spectra supplemented by additional optical spectra
to explore the spectral features associated with the central binary and 
its immediate circumstellar environment.  Our primary goal is to determine 
how the binary ejects the gas that ultimately flows into the dense outer double-toroidal nebula.
We describe our observations and data reduction methods in \S2 and 
we present in \S3 examples of the orbital phase-related spectral variations we observed.
The radial velocity curve and a new orbital 
solution for the supergiant mass donor are described in \S4. 
Then in \S5 we review the spectral clues about the nature of the enshrouded mass gainer,  
and we present a preliminary radial velocity curve and orbital solution for the massive companion.
Both radial velocity curves are used in \S6 to make a Doppler tomographic reconstruction of 
the optical spectra of the individual components.
We describe in \S7 a model for the mass outflows in RY~Scuti that is based upon
recent hydrodynamical simulations and that explains many of the observed spectral variations. 
Our results are summarized in \S8.


\section{Observations and Reductions}

Our analysis is based upon spectra collected from three telescopes. 
The highest dispersion spectra (17 in number) were obtained in 1999 with the 
Fiber-fed Extended Range Optical Spectrograph 
(FEROS) mounted on the 1.52~m telescope of the European Southern 
Observatory (ESO) at La Silla, Chile (see \citealt{smi02}).
Also in 1999, we obtained 40 moderate dispersion spectra (in the red region 
surrounding H$\alpha$) using the Kitt Peak National Observatory 
coud\'{e} feed 0.9~m telescope.  Finally, in 2004 we used the CTIO 1.5~m telescope 
and Cassegrain spectrograph to obtain ten blue spectra of moderate resolution
covering one orbital period.  Table~\ref{ScopeInfo} contains run number, dates, 
spectral coverage, spectral resolving power, number of spectra, 
telescope, spectrograph grating, and CCD detector used in each case.
Exposure times were generally limited to 30 minutes or less. 
Each set of observations was accompanied by numerous bias, 
flat-field, and ThAr comparison lamp calibration frames.  Furthermore, we 
obtained multiple spectra each night of the rapidly rotating A-type 
star $\zeta$ Aql for removal of atmospheric water vapor and O$_2$ bands
in the red spectra made with the KPNO coud\'{e} feed.

\placetable{ScopeInfo}  

The spectra from all the telescopes were extracted and calibrated 
using standard routines in IRAF\footnotemark
\footnotetext{IRAF is distributed by the National Optical Astronomical
Observatory, which is operated by the Association of Universities for
Research in Astronomy, Inc. (AURA), under cooperative agreement with the
National Science Foundation.}.
All the spectra were rectified to a unit continuum by fitting 
line-free regions.  We removed the atmospheric lines from the red
coud\'{e} feed spectra by creating a library of $\zeta$ Aql spectra from each run, 
removing the broad stellar features from these, and then dividing each target 
spectrum by the modified atmospheric spectrum that most closely matched the 
target spectrum in a selected region dominated by atmospheric absorptions. 
In a few cases this resulted in the introduction of spectral discontinuities 
near the atmospheric telluric lines, and these were excised by linear interpolation.  
We did not attempt any removal of atmospheric lines for the FEROS
spectra, but some problem sections in the echelle-overlap regions were excised 
via linear interpolation.  The spectra were then transformed to a common 
heliocentric wavelength grid for each of the FEROS, KPNO coud\'{e} feed, and
CTIO 1.5~m runs.   We show several examples of the final spectra in the next section.


\section{Spectral Variations with Orbital Phase}

The optical spectrum of RY Scuti is very complex and time variable, 
and in this section we describe the appearance and orbital 
phase-related variations of several representative line features. 
A depiction of the time-average of all the FEROS spectra in the range 
from 3600 to 9200 \AA\ appears in \citet{smi02} (their Fig.~16). 
This illustration is ideal for identifying the many  
sharp, quadruple-peaked, emission lines formed in the surrounding 
double-toroidal nebula, but photospheric lines appear broad and shallow 
as the average was made over the full range of orbital Doppler shifts. 
A large fraction of the circumstellar emission lines are species that
do not have a stellar absorption counterpart (such as [\ion{Fe}{3}] and 
many other forbidden lines), but there are several emission lines that 
are superimposed upon important stellar spectral features 
(for example, in the H Balmer and \ion{He}{1} lines). 
Since the nebular features are analyzed in great detail elsewhere
\citep{golsk92,skwe93,smi02}, we will not discuss them here.
We will focus on those spectral features formed in the photospheres of 
the stellar components and in the rapidly moving gas immediately surrounding them.  
Here we introduce the different kinds of patterns of variability observed, and 
in the following sections we analyze in detail the Doppler velocity shifts
related to features associated with the supergiant (\S4) and its massive companion (\S5). 

We begin with some examples from the high resolution FEROS spectra. 
The \ion{Si}{4} $\lambda4088$ absorption feature is one of only a  
small number of spectral lines formed in the photosphere of the supergiant 
that is not filled in or affected by nebular emission.
We show the orbital phase variations of this line (and the 
nearby \ion{N}{3} $\lambda 4097$ and H$\delta$ lines) in Figure~\ref{SiIV4088}
as a function of heliocentric radial velocity for \ion{Si}{4} $\lambda4088$. 
In this and the next figures we adopt the orbital period from \citet{kre04}
of $P=11.12445$~d and the epoch of phase zero as the supergiant 
superior conjunction at $T_{SC}={\rm HJD}~2,451,396.71$ (derived in \S4).
The upper portion of Figure~\ref{SiIV4088} shows the \ion{Si}{4} profiles with 
their continua aligned with the phase of observation (increasing downwards)
while the lower portion shows the spectra as a gray-scale image 
interpolated in phase and velocity.  Features moving with the radial 
velocity curve of the supergiant will have a characteristic ``S'' shape
in this image.  The break in the continuity of the ``S'' curve near 
phase $\phi=0.35$ is due to the unfortunate gap in our phase coverage near 
there and to the simplicity of the interpolation scheme.   
The \ion{Si}{4} $\lambda4088$ feature appears to be useful for the 
radial velocity measurement of the supergiant, although we find that 
the depth of the line varies with phase, weakening at $\phi=0.0$ and 
strengthening at $\phi=0.5$.  There is no obvious evidence of a reverse ``S''
feature that would correspond to the motion of the massive companion
(although a weak feature is present; \S6). 

\placefigure{SiIV4088}  
 
The orbital variations in two helium lines, the \ion{He}{1} $\lambda4387$ singlet 
and the \ion{He}{1} $\lambda4471$ triplet, are illustrated in 
Figures~\ref{HeI4387} and \ref{HeI4471}, respectively.  
Both features show sharp nebular emission peaks from the surrounding 
double-torus \citep{smi02} superimposed on the stellar absorption feature.  
Both lines show an absorption component that follows the radial velocity curve of the supergiant. 
However, the \ion{He}{1} $\lambda4387$ feature strengthens at both conjunctions 
and almost disappears at phase $\phi=0.25$.  It also shows a blueshifted absorption 
feature that appears at $\phi=0.5$ and lasts for nearly half the orbit.
Other \ion{He}{1} singlets ($\lambda\lambda4009,4143,4921$) exhibit the same behavior.
On the other hand, the \ion{He}{1} $\lambda4471$ triplet displays a much 
stronger blueshifted component between $\phi=0.5 - 0.8$ that develops
into a very sharp and blueshifted ($\approx -200$ km~s$^{-1}$) absorption
line that lasts for greater than half the orbit.  The other \ion{He}{1} triplets
($\lambda\lambda4026,4713$) also show these same features. 

\placefigure{HeI4387}  

\placefigure{HeI4471}  

Figure~\ref{SiIII4552} shows the orbital variations in the triplet 
\ion{Si}{3} $\lambda\lambda4552,4567,4574$ in the velocity frame of 
\ion{Si}{3} $\lambda4552$.  The supergiant component appears to 
be present and undergoes the same kind of strengthening at 
conjunctions seen in the \ion{He}{1} lines.   However, each of 
the triplet members also show a broad shallow feature that moves 
in the manner expected for the massive companion.  Thus, our observations
confirm the detection of this second component that 
was discovered by \citet{skul92}. 

\placefigure{SiIII4552}  

There are two very broad emission lines present that are important to the study of 
this binary system, H$\alpha$ $\lambda6563$ and \ion{He}{2} $\lambda4686$ \citep{sah02}.
The H$\alpha$ feature consists of a broad emission feature that 
is spatially coincident with the central binary plus a strong but narrow
component formed in the double-toroidal nebula \citep{smi02}.  Therefore,
in order to isolate the emission component near the binary, we had to remove the nebular
components of H$\alpha$ and the nearby [\ion{N}{2}] $\lambda\lambda6548,6583$ lines.  
This removal process was done by scaling the [\ion{N}{2}] $\lambda6583$ line 
in each spectrum to the 
appropriate size (of either H$\alpha$ or [\ion{N}{2}] $\lambda6548$) using the
equivalent widths of these lines from \citet{smi02}, 
shifting the rescaled line to the location of the line to be removed, 
and then subtracting it from the spectrum.  This process was done interactively
and included small adjustments in the scaling and shifting parameters to 
optimize the subtraction.   
The resulting subtracted H$\alpha$ profiles based upon 
the large set of KPNO coud\'{e} feed spectra are illustrated as a function
of orbital phase in the left panel of Figure~\ref{HaCoude}.  
We also show the one spatially resolved {\it HST} spectrum of the central 
binary from the work of \citet{smi02} (which has no nebular emission present) 
that verifies that our line subtraction technique creates difference profiles 
with the appropriate shape. 
The emission strength appears to be much stronger at the conjunction phases, 
but this is due mainly to the drop in the continuum flux at those eclipse phases 
and our normalization of the emission strength to this varying continuum
level.  The right panel in Figure~\ref{HaCoude} shows a representation of the
H$\alpha$ profiles relative to a constant flux continuum we made by rescaling 
the emission flux by a factor
$$f/f({\rm quadrature})=10^{-0.4(V-9.03)}$$
where $V$ is the $V$-band magnitude at the orbital phase of
observation (found by interpolation in the light curve data from
\citealt{djur01}) and $V=9.03$ corresponds to the maximum brightness
of the system at quadrature phases.  The H$\alpha$ emission feature
appears broad (spanning over 1000 km~s$^-{1}$) and approximately
constant in strength and position in this version.  
There is often a weak, blueshifted absorption component present 
(with a radial velocity of $\approx -150$ km~s$^{-1}$) that gives the profile 
a P~Cygni appearance.  
Similar results were seen in the H$\alpha$ difference profiles formed from the smaller 
set of FEROS spectra. 

\placefigure{HaCoude}  

Finally we show in Figure~\ref{HeII4686} the orbital phase variations in the weak
\ion{He}{2} $\lambda4686$ emission line observed in the FEROS spectra that we
smoothed with a Gaussian of FWHM = 45 km~s$^{-1}$ to improve the otherwise noisy appearance.
This feature is found only in very hot plasmas, and there is no corresponding 
nebular feature in this case.  The \ion{He}{2} $\lambda4686$ emission is almost 
as broad as the net H$\alpha$ emission, but is significantly weaker.  
The wings of the line appear to exhibit a slight anti-phase motion that 
was documented by \citet{sah02}.  

\placefigure{HeII4686}  


\section{Radial Velocity Curve of the Supergiant}

The radial velocity shifts of the supergiant are readily apparent 
in many lines, and given the quality and number of the new spectra
we decided to measure the radial velocities and reassess the orbital elements. 
We measured relative radial velocities by cross-correlating 
each spectrum with a template spectrum.  
For the FEROS and CTIO spectra, this template spectrum was generated 
from the non-LTE, line-blanketed model atmosphere and synthetic spectra grid 
from \citet{lan03} using $T_{\rm eff} = 30$ kK and $\log g = 3.0$, 
which are appropriate values for an O9.7 Ib star \citep{her95}.
This template was rotationally broadened (\S5) and also smoothed to the 
instrumental resolution of the FEROS or CTIO spectra.
First, we removed certain interstellar features by forming an average 
interstellar spectrum from the mean of the entire set and then 
dividing each spectrum by the average interstellar spectrum.
Because the spectrum of RY~Scuti contains so many stationary nebular lines 
and possible features from the massive companion, we restricted 
the wavelength range for the cross-correlation to regions surrounding
a set of absorption lines that appeared to be free of line blending and 
that were clearly visible throughout the orbit.  The main features
in the selected regions are summarized in Table~\ref{linelist}.  

\placetable{tab2}  

For the KPNO coud\'{e} feed spectra, we selected one spectrum from the 
set of 40 to serve as the template since the Lanz \& Hubeny models did not
match the emission lines we measured (Table~\ref{linelist}).  
This spectrum (made on
HJD 2451425.784) has relatively high signal-to-noise, exhibits 
well-defined spectral features, and was made near quadrature 
when the supergiant Doppler shifts are large and the features reasonably 
well separated from any component from the massive companion.    
Once again, we first removed interstellar 
features then selected regions free from nebular lines for the
cross-correlation.  Once we performed the cross-correlation, we determined 
the absolute radial velocity of the template spectrum by fitting the core 
of the emission lines (\ion{N}{2} $\lambda6610$ and \ion{Si}{4} 
$\lambda6701$) with a parabola and determined the shift from the rest 
wavelengths. 
We added the radial velocity of the template to all the relative velocities
to place them on an absolute scale. 

Since the velocities from the different runs are based upon measurements
of different lines in the template and individual spectra, we anticipated 
that there might be systematic differences in the zero-points of each set. 
We began by making independent circular orbital fits for each set, and indeed
we found the systemic velocity $\gamma$ for the FEROS set was offset by 
$-20.6$ km~s$^{-1}$ from the resulting systemic velocities for the 
KPNO and CTIO sets.  Since the latter were closer to the nebular systemic
velocity ($20\pm 3$ km~s$^{-1}$; \citealt{smi02}) and based upon more
observations, we arbitrarily adjusted the FEROS measurements by adding
$20.6$ km~s$^{-1}$ to bring them into consistency with the other measurements. 
Note that this decision adds an additional uncertainty to the real error
in the final determination of the systemic velocity 
but otherwise has no effect on the other orbital elements.  Our final radial 
velocities are collected in Table~\ref{SGrv} that lists the date of 
observation, orbital phase (see below), radial velocity, calculated error
in radial velocity, observed minus calculated residual,
and the telescope where the spectrum was made (Table~\ref{ScopeInfo}).
The quoted radial velocity error for the blue spectra is
the standard deviation of the cross-correlation measurements
from the three line regions.  We measured only two line regions
in the red spectra, so the velocity error is estimated as the
larger of the difference between the two measurements or
the mean value of $|V_i-V_{i+1}|/\sqrt{2}$ from closely spaced
pairs of observations.

\placetable{SGrv} 

We determined revised orbital elements using the the nonlinear,
least-squares, fitting code of \citet{morb74}. 
Since the errors associated with each run are different, 
we assigned each measurement a weight proportional to the inverse square
of the measurement error.
The orbital period was fixed at $P=11.12445$ days as found by \citet{kre04}
from contemporary eclipse timing observations over a long time base.  
We made both eccentric and circular orbital solutions, but we think 
the eccentric solution, while formally statistically significant according 
to the test of \citet{lucy71}, is probably spurious.  
\citet{har87} found that gas streams and circumstellar matter (both present in
RY~Scuti) can distort spectroscopic features in 
mass-transferring binaries.  Such distortions can lead to skewed radial 
velocity results and thus artificial eccentricities \citep{luc05}. 
A zero eccentricity is consistent with predictions for Roche lobe
overflow systems where the tidal effects are expected to 
circularize the orbit and synchronize the rotational and orbital periods. 
The final orbital elements from both the circular and eccentric solutions 
are presented in Table~\ref{OrbSol} 
together with solutions from \citet{sah02} and \citet{skul92}. 
The epoch $T_{SC}$ refers to the time of supergiant superior 
conjunction (close to the time of photometric minimum light)
while $T$ is the epoch of periastron in the eccentric solution. 
The radial velocities and orbital velocity curves are plotted in 
Figure~\ref{RV-all}.
Our results are in reasonable agreement with earlier determinations
of the elements, with the possible exception of the larger non-zero 
eccentricity found by \citet{sah02}.  We suspect the 
difference is due to the lack of orbital phase coverage in the 
observations of \citet{sah02} in the ascending portion of the velocity curve.  

\placetable{OrbSol} 

\placefigure{RV-all} 


\section{Nature of the Massive Companion}

Very little is known about the massive companion because it is 
difficult to find its associated features in the spectrum 
(probably because the massive companion is enshrouded in  
a thick accretion disk; \S7).   We can make an approximate 
estimate for the expected semiamplitude of motion for the 
massive companion using a geometrical argument to find the mass 
ratio, $q=M_{MC}/M_{SG}$.  Let us assume the supergiant 
is filling its Roche lobe and is synchronously rotating
(reasonable in the case of RY~Scuti where we see so much 
spectroscopic evidence of active mass transfer). 
\citet{gies86} showed that for such Roche-filling stars,
the projected rotational velocity $V\sin i$ provides a 
measure of the stellar radius while the semiamplitude $K$ 
is related to the mass ratio and semimajor axis $a$, and the 
ratio of these quantities is a function that solely depends 
on the mass ratio, 
\begin{eqnarray*}
V \sin i / K & = & {{\Omega R_{SG} \sin i}\over{\Omega a_{SG} \sin i}} \\
& = & \left(1+{M_{SG}\over M_{MC}}\right) {R_{SG}\over a} = \left(1+Q\right) f(Q)
\end{eqnarray*}
where $\Omega$ is the orbital angular rotation speed, 
$Q=1/q=M_{SG}/M_{MC}$, and $f(Q)$ is the fractional 
Roche filling radius \citep{egg83}.  Thus, a measurement of the
supergiant's projected rotational velocity leads directly to 
an estimate of the mass ratio.  We show below that
this geometrical mass ratio estimate is consistent with
the mass ratio derived from orbital velocity measurements for
both components. 

We determined $V\sin i$ for the supergiant through a comparison 
of the photospheric line width with models for a grid of projected rotational velocity.  
For this purpose, we selected the \ion{Si}{4} $\lambda4088$ line since 
it is free of nebular emission and line blending with other features 
and its shape is dominated by rotational broadening. 
We measured the FWHM of \ion{Si}{4} $\lambda4088$ in 
three of the FEROS spectra obtained near supergiant maximum velocity 
(to avoid the unusual strengthening seen at conjunctions and 
any line blending problems with absorption from the massive companion). 
We then used the model profiles from the non-LTE, line blanketed 
models of \citet{lan03} that were rotationally broadened by
a simple convolution of the zero-rotation model profiles with a rotational
broadening function \citep{gray92} using a linear limb-darkening coefficient
from the tables of \citet{wr85}.  From the Lanz \& Hubeny grid, we selected
a model profile for 
$T_{\rm eff} =30$~kK and $\log g = 3.0$ and adopted a linear 
limb darkening coefficient at this wavelength of $\epsilon = 0.38$. 
Finally, the resulting models were convolved with an instrumental 
broadening function to match the resolution of the FEROS spectra. 
The projected rotational velocity for the supergiant derived from 
the resulting $(V\sin i, FWHM)$ relation is $V\sin i = 80 \pm 5$ km~s$^{-1}$.  
The error represents the standard deviation of $V\sin i$ as derived
from the three measurements of line width and does not account for
any systematic errors associated with the choice of model atmosphere
parameters.

We caution that this is actually an upper limit for $V\sin i$ 
since the line may also be broadened by macroturbulence 
in the stellar atmosphere \citep{rya02}.  If we include an estimate
of $V_{macro}=30$ km~s$^{-1}$ 
(a mid-range value for B-supergiants; \citealt{rya02}), then we find 
$V_{eq} \sin i = 74^{+7}_{-13}$ km~s$^{-1}$ (where the error range 
accounts for the derived error in $V\sin i$ and a range in macroturbulent 
velocity from 10 to 50 km~s$^{-1}$).
Thus, using the $V\sin i / K$ method described above, 
we find an estimate of the mass ratio, $q=4.9^{+2.9}_{-0.9}$.  
We can use this mass ratio estimate to offer some guidance about
the expected orbital semiamplitude of the massive companion, 
$K_{MC}=K_{SG}/q= 31 - 61$ km~s$^{-1}$.

Previous researchers \citep{pop43,cowhut76,skul92,sah02} have associated 
a variety of spectral features with the massive companion, and we 
inspected the orbital variations of all these proposed lines in the FEROS 
spectra (examples shown in Fig.~\ref{SiIV4088} -- \ref{SiIII4552}).  
Many of these lines have complicating factors related to nebular emission 
and blending with other features.  For example, a number of the \ion{He}{1} 
lines appear to show a blueshifted absorption feature in the phase range 
$\phi=0.6-0.9$ where we would expect to find the absorption component
from the massive companion, but these blueshifted features first appear
at $\phi=0.5$ when any gas leaving the supergiant in the L2 region 
would also appear blueshifted (\S7).  Consequently, an identification
of these features with the massive companion is ambiguous.  

Only three of the proposed spectral features for the massive companion
clearly fit the mass ratio argument described above and display the 
expected anti-phase velocity shifts:
H$\alpha$, \ion{He}{2} $\lambda4686$ and \ion{Si}{3} $\lambda4552$.  
Here we present radial velocity measurements for each of these.  
The H$\alpha$ emission profile is not Gaussian in shape (see Fig.~\ref{HaCoude}) 
and therefore could not be measured as such.  Instead we determined the radial 
velocity of the line wings (which represent the fastest-moving gas 
and are unaffected by nebular emission) based upon a 
bisector position found using the method of \citet*{sha86}.  
We sampled the line wings using oppositely signed Gaussian 
functions and determined the mid-point position between the wings by 
cross-correlating these Gaussians with the profile.
We used Gaussian functions with FWHM = 137 km~s$^{-1}$ at sample 
positions in the wings of $\pm382$ km~s$^{-1}$ for spectra from both
the coud\'{e} feed and the FEROS instrument.  
An inspection of radial velocity measurements closely spaced
in time indicates an average error of 1.7 km~s$^{-1}$.
Our results are given in column~3 of Table~\ref{Hawings}, and the values are
plotted as a function of orbital phase in Figure~\ref{RV-all}. 

\placetable{Hawings} 

The \ion{He}{2} $\lambda 4686$ emission profile is also broad 
(FWHM $\approx 850$ km~s$^{-1}$) and not Gaussian-shaped.
Therefore, we used spectral templates and cross-correlation 
functions to measure radial velocities.  We formed templates 
from the averages of the entire run for both the FEROS and 
CTIO data sets since the \ion{He}{2} $\lambda 4686$ emission 
strength may have changed between 1999 (FEROS) and 2004 (CTIO).
We then transformed the relative velocities to an absolute scale 
by adding the radial velocity derived by fitting each template 
with a broad Gaussian.  Our results are given in column~4 of 
Table~\ref{MCfeatures} and are plotted in Figure~\ref{RV-all}.
Based on visual estimates of the goodness of the fit of this 
very broad line, we estimate errors in the velocities are 
$\approx$ 50 km~s$^{-1}$.

\placetable{MCfeatures} 

We also measured equivalent widths for both these emission lines
by a direct numerical integration (over the range $6536.3 - 6595.0$~\AA\
for H$\alpha$ and $4675.3 - 4694.5$~\AA\ for \ion{He}{2} $\lambda 4686$). 
Our results for H$\alpha$ are given in column~4 of Table~\ref{Hawings} and
for \ion{He}{2} are given in column~5 of Table~\ref{MCfeatures}. 
Errors in H$\alpha$ measurements are $\approx$0.5~\AA\ based upon examining 
results closely spaced in time.  The \ion{He}{2} equivalent width errors 
are 0.02~\AA\ for the FEROS data but are about three times worse for the 
CTIO data (due to the lower resolution of the CTIO spectra).
The orbital variations of these equivalent widths are discussed below (\S7).

The absorption line \ion{Si}{3} $\lambda4552$ displays narrow and broad components 
that move with the expected orbital Doppler shifts of the supergiant 
and massive component, respectively (Fig.~\ref{SiIII4552}).  In order to avoid any 
line blending problems between these components, we selected a subset of 
FEROS and CTIO spectra observed near the quadrature phases where the 
the broad component could be measured unambiguously, and we made 
Gaussian fits to determine the radial velocities.  
Using a visual estimate of goodness of fit to this broad, shallow line, 
we estimate errors in these velocities are $\approx$ 30 km~s$^{-1}$.
Our results are given in column~3 of Table~\ref{MCfeatures} and are 
plotted in Figure~\ref{RV-all}.

The plots of the radial velocities in Figure~\ref{RV-all} show that 
all three features exhibit the anti-phase motion expected for the massive companion.
We made circular fits of each set by fixing the orbital period and 
epoch from the solution for the supergiant (Table~\ref{OrbSol}) and then solving for 
the semiamplitude $K_{MC}$ and systemic velocity $\gamma_{MC}$.   
The results are listed in Table~\ref{MCRVfits}.  The semiamplitudes 
for each feature are within the expected range, however, the values 
for the two emission lines are smaller than that for the 
\ion{Si}{3} $\lambda 4552$ absorption line.  Furthermore, the 
systemic velocities for the two emission lines are larger than 
that found for the supergiant and the nebular emission. 
We suspect this is due to the P~Cygni shape of both features 
that will bias an emission line velocity measurement to a larger value. 
We argue below (\S7) that these emission features are probably 
formed in gas flows and are thus probably less representative of 
the orbital motion of the massive companion.   

\placetable{MCRVfits} 

Thus, if we adopt the \ion{Si}{3} $\lambda 4552$ velocity fits as 
the most representative of the massive companion, we can make a 
double-lined solution of the spectroscopic masses, $M\sin^3 i$,  
using eq. 2.52 in \citet{hil01},
$$M_{1,2} \sin^3 i=(1.0361\times10^{-7})(1-e^2)^{3/2} (K_1+K_2)^2 K_{2,1} P~M_\odot,$$
where $K$ is the semiamplitude in km~s$^{-1}$ and $P$ is the period in days.
Furthermore, if we set the orbital inclination equal to that for the 
double-toroidal nebula, then $i=75\fdg6 \pm 1\fdg7$ \citep{smi99} 
(in substantial agreement with the light curve solutions; 
\citealt{mil81,djur01}), and we can estimate the component masses directly.  
We find the masses of the components are
$M_{SG} = 7.1 \pm 1.2 M_\odot$ and $M_{MC}=30.0 \pm 2.1 M_\odot$.  
The former is very low for a normal O-supergiant \citep*{mar05}, 
but we should bear in mind this star is in the process of losing 
its outer envelope to the massive companion and therefore its properties 
will be very different from those of a single star of comparable 
temperature and gravity.  The mass of the massive companion is like that 
of a hot O-dwarf or an evolved B-supergiant and we will explore these 
and other possibilities in the next sections.   Our mass results are 
quite similar to those first obtained by \citet{skul92} but are 
lower than those determined by \citet{sah02} (given in Table~\ref{OrbSol}).


\section{Tomographic Reconstruction of Spectra}

Once the orbital solution for RY Scuti was found, we used a tomographic reconstruction 
technique \citep{bag94} with the FEROS data to separate the individual 
spectra of the components.  Tomographic reconstruction is an 
iterative scheme that uses the combined spectra and their 
associated radial velocities to determine the appearance of each star's spectrum. 
RY~Scuti presents a special difficulty due to the stationary sharp nebular features.  
Therefore, before reconstruction, all nebular features 
listed by \citet{smi02} were excised via linear 
interpolation.  We also removed the interstellar lines from each spectrum prior to 
reconstruction to avoid creating spurious reconstructed features in their vicinity.
The reconstruction was based upon a subset of seven FEROS spectra that 
were obtained near the velocity extrema at the quadrature phases so that
we might avoid introducing artifacts due to the line strengthening at the conjunctions
and any eclipse effects. 
Figures \ref{tomog39} and \ref{tomog44} show plots of the reconstructed spectra 
in two different regions along with identifications of the principal lines.  
Also plotted are two comparison spectra from the
\citet{val04} {\it Indo-U.S.\ Library of Coud\'{e} Feed Stellar Spectra} that have a lower 
resolving power ($R\approx 3600$) than that of the FEROS spectra ($R\approx 48000$).

\placefigure{tomog39}  

\placefigure{tomog44}  

The supergiant has a classification of O9.7~Ibpe~var \citep{wal82} 
and we see the spectrum of the normal single star, HD~188209
(O9.5~Iab; \citealt{wal76}), provides a reasonably good match.  
The \ion{He}{2} $\lambda\lambda 4199, 4541, 4686$ features appear to 
be slightly weaker in the RY~Scuti supergiant, which is consistent
with its subtype difference in spectral type.  We used the ratios 
of the equivalent widths of several weaker \ion{He}{1} lines for 
the RY~Scuti supergiant and HD~188209 spectra to set the monochromatic 
flux ratio for the RY~Scuti binary components, and we find the 
supergiant contributes $\approx 57 \pm 13 \%$  of the total flux.  The error is
based on scatter between the results from different spectral features.
The reconstructed spectra are plotted in Figures \ref{tomog39} and \ref{tomog44} 
normalized to their respective continuum fluxes, so individual lines appear 
deeper than in the composite spectra where each is diluted by the flux of the
other star.
There are several obvious differences between the supergiant component
and single star spectrum in some of the stronger lines such as 
H$\epsilon$ $\lambda3970$ and H$\delta$ $\lambda4101$ 
but we do not ascribe any particular significance to these as
such features are clearly distorted by emission from circumstellar gas. 
On the other hand, there does appear to be a significant weakness in 
the C lines and an enhancement in the N lines in the spectrum of the 
RY~Scuti supergiant compared to that of HD~188209.  For example, 
the \ion{C}{3} $\lambda\lambda 4070,4647$ lines that are present in the spectrum 
of HD~188209 are conspicuously absent in the RY~Scuti supergiant's spectrum. 
This suggests the atmosphere of the supergiant is enhanced 
with CNO-processed gas (as is the surrounding nebula; \citealt{smi02}). 
Perhaps this is not surprising given that a large portion of the supergiant's
mass must have already been lost to reveal gas from deeper layers closer
to the source of core H-fusion. 
This result suggests that the bright O9.7 supergiant is the ultimate
source of the processed gas that now forms the outer double-toroidal nebula.

\citet{skul92} suggested the spectral features moving with Doppler
shifts of the massive companion have the appearance of a B2 star. 
We find the lines in the reconstruction for the orbital shifts
of the massive companion more closely resemble that of the B0.5~Ia star, 
HD~185859 \citep*{mor55}.  However, there are several features in 
the reconstruction that make us caution against assuming 
this spectrum forms in the photosphere of a supergiant star. 
First, the luminosity sensitive lines of \ion{Si}{4} $\lambda\lambda 4088,4116$
and \ion{Si}{3} $\lambda\lambda 4552,4567,4574$ appear very strong 
in the reconstructed spectrum relative to those in HD~185859 
which would imply the lines form in a very low density plasma 
(like those in the photospheres of the most luminous stars; \citealt{wal90}). 
Secondly, if we suppose that the gravity of a B0.5~Ia star is $\log g =3.0$
(see the case of the like star $\kappa$~Ori; \citealt*{mce99}),
then using our derived mass, the massive companion
star would have a radius of $\approx 29 R_\odot$, which is 83\% of its 
Roche radius.  A radius this big would have two observational consequences:
first, the flux from this component would be twice the flux from the
bright 09.7 supergiant, and second, a star this large would cause the
spectral lines of the bright 09.7 supergiant to weaken or disappear during
primary eclipse ($\phi=0.0$).  Neither of these two predictions are seen: 
the massive companion flux is less than that of the supergiant and the 
supergiant's absorption lines do not weaken at primary eclipse 
(Fig.~\ref{SiIV4088} -- \ref{SiIII4552}).
Also, we estimate the projected rotational velocity of the massive 
companion spectrum  by measuring the
widths of the \ion{Si}{3} $\lambda 4552$ feature in five FEROS
spectra where the feature was well separated from any component of
the bright supergiant.  We used these widths to obtain
$V\sin i = 194 \pm 18$ km~s$^{-1}$ in the same way
as described above for the primary (\S5) by
comparing them with rotationally broadened model spectra for
$T_{\rm eff} = 27500$~K, $\log g = 3.0$, and a limb darkening
coefficient $\epsilon = 0.34$.   This result is unreasonably large since
\citet{rya02} find that no early B-type supergiant has a $V\sin i$
greater than 60 km~s$^{-1}$.
Therefore, we suggest these characteristics are best explained if we place the 
origin of the spectrum in the photosphere of a thick accretion disk
surrounding the massive companion rather than in the photosphere of the star itself 
and we explore this idea further in the next section.  Note that the
very broad \ion{He}{2} $\lambda4686$ emission is assigned to the massive companion
in the reconstruction, but we argue below that its site of origin may not be 
exactly co-spatial with that of the absorption line spectrum.


\section{Mass Outflows in RY Scuti}

Many of the spectral features discussed above originate in gas outflows from 
the binary, and in this section we explore their sites of origin with 
reference to recent hydrodynamical simulations of gas flows in Roche lobe 
overflow binaries.  
The luminous supergiant in RY~Scuti can potentially be losing mass
by a radiatively driven wind and/or tidal streams along the axis
joining the stars.  \citet{fri82} show that a continuum of
states can exist between these idealized cases.  When a luminous
star has a radius that is significantly smaller than the critical Roche lobe,
the wind will be fast and spherically symmetric (as for single stars).
As the ratio of stellar to Roche radius increases, the star becomes tidally
distorted and its wind becomes asymmetric with a higher mass loss
rate and slower outflow along the axis joining the stars
(a ``focused wind'').  Finally, once the star essentially attains
a size filling the Roche surface, most of the mass loss will occur in
slow outflows directed along the axis joining the stars.   All of the
photometric light curve studies confirm the supergiant is very tidally
distorted and must therefore be close to Roche filling.  Thus, we
will explore the consequences of tidally formed outflows in
reviewing the observational clues about the mass loss.

The discussion begins with the dimensions of the system, and in 
Figure~\ref{cartoon} we present a cartoon illustration of the cross section of 
the system as viewed from above the orbital plane.  The direction of the 
observer at different orbital phases is indicated by numbers in the 
periphery of the diagram.  
As before, we assume the supergiant fills its critical Roche surface and 
thus has an equivalent volume radius of approximately 
$R_{SG}=18 R_\odot$ \citep{egg83}.
The supergiant is depicted on the left side of Figure~\ref{cartoon}.  
If we assume the massive companion is a main sequence star, then for its 
derived mass we expect it is an O6.5~V star with a polar radius of 
approximately $9.6 R_\odot$ (see Table~4 in \citealt{mar05}).  
The equatorial radius could be significantly larger if the star is rapidly 
rotating, but we assume it is a spherical star in its depiction on the 
right hand side of Figure~\ref{cartoon}.  
Either way, the radius of the massive companion is probably 
significantly less than its critical Roche radius, and therefore we 
begin our analysis assuming that the system is semi-detached.  

\placefigure{cartoon}  

The kinds of gas outflows expected in systems like RY~Scuti 
are explored in a series of recent papers by \citet{naz03,na06a,na06b}.  
They present hydrodynamical simulations
in two and three-dimensions to model the case of the interacting 
binary $\beta$~Lyr \citep{har96,har02}.  RY~Scuti and $\beta$~Lyr
share many of the properties common to the W Serpentis class, 
and we can obtain considerable insight 
about the gas flows in RY~Scuti from these papers (although we 
caution that the stars in RY~Scuti are both hotter and more massive than
those in $\beta$~Lyr).  These numerical models assume the mass 
donor is a low gravity object that fills its Roche surface, and the 
authors use the models to follow the time evolution of the gas flows 
until they reach a semi-static configuration.  There are several 
features of these models that are relevant to the case at hand. 
First, Nazarenko \& Glazunova find that the donor loses mass through 
both the inner L1 region and the outer L2 region (facing away from the
companion).  Because the gas is so loosely bound to the donor, 
these gas streams are not compressed into a classical stream but 
instead fan out across a significant portion of the stellar surface 
facing towards and away from the companion.  The approximate ranges of 
these outflows are indicated by the dashed lines to the left and right 
of the supergiant in Figure~\ref{cartoon} and are based upon simple 
ballistic trajectories.  
The thicker line to the right of the supergiant indicates the classical
L1 gas trajectory \citep{lub75}.  The models also show an 
accretion disk does form but that gas may be lost from the binary near the outer L3 
point (to the right of the massive companion in Fig.~\ref{cartoon}).  We have 
simply illustrated the disk in Figure~\ref{cartoon} assuming it extends to
$80\%$ of the Roche radius of the massive companion ($R_{Roche} \approx 35 R_\odot$).  
The three-dimensional models \citep{na06a,na06b} show the inner portions of 
the disk attain a significant height above and below the orbital plane, and
therefore the gas structure is referred to as a torus rather than a disk.  
If the massive companion is a main sequence star, then the height of the 
torus may well exceed the stellar radius and block it from view for 
all but extremely low orbital inclinations.  Finally, \citet{na06b} 
consider the effect that a stellar wind from the mass gainer will have 
on the surrounding torus.  They find the wind breaks out in 
bipolar outflows and that the hottest gas temperatures occur where 
the wind strikes the inner torus.  Such a bipolar outflow may explain 
the elongated radio emission surrounding $\beta$~Lyr \citep{uma00}.
If the massive companion in RY~Scuti is an O6.5~V star, then we expect 
it will be hot and luminous (perhaps exceeding the luminosity of 
normal dwarfs due to the addition of accretion luminosity) and 
thus it will be a source of a significant radiatively-driven wind. 

We will use the framework of gas features illustrated in Figure~\ref{cartoon} 
to offer a plausible origin for some of the spectral features discussed 
in earlier sections.  We begin with the expected outflows from the 
supergiant towards L1 and L2.  Most of the absorption 
lines associated with the supergiant appear to strengthen considerably 
in depth at the conjunction phases and this change is particularly 
striking in the case of the \ion{He}{1} lines illustrated in 
Figures \ref{HeI4387} and \ref{HeI4471}.  
The absorption preferentially strengthens on the blue side of the profile, 
and in the case of supergiant inferior conjunction ($\phi=0.5$), remains
blueshifted for a while longer.  For triplet transitions like \ion{He}{1}
$\lambda 4471$ (Fig.~\ref{HeI4471}), the blueshifted component transforms from 
a broad feature into a narrow, almost nebular absorption line that lasts
for more than half the orbit.  We suggest these changes are best explained 
by absorption from the outflows from the supergiant that emerge in 
the L1 and L2 directions.  The blueshifts we observe with projected speeds of
$\approx$ 200 km~s$^{-1}$
are consistent with 
an outflow towards L1 (at $\phi=0.0$) and towards L2 (at $\phi=0.5$). 
The disappearance of the broad absorption shortly after $\phi=0.0$ is 
expected since this flow terminates only a short distance from the 
supergiant where the gas strikes the accretion disk.  On the other hand, 
the gas leaving the L2 region forms a long trailing plume that may occult 
the supergiant for a considerable time following $\phi=0.5$ and thus may
account for the presence of blueshifted absorption until $\phi\approx0.8$.
The appearance of the narrow blueshifted absorption in the \ion{He}{1}
triplets may be due to low optical depth gas associated with the L2 outflow 
that extends well above and below the orbital plane. 

The gas that enters the accretion disk from the L1 outflow is probably very
dense and extended from the orbital plane.  In such a situation, the 
massive companion may be wholly obscured from our point of view and 
its photospheric spectrum will be invisible.  However, we noted in \S5 
and \S6 that there are a number of spectral features that do display  
radial velocity patterns opposite to those of the supergiant and 
may plausibly be associated with the immediate environment of the 
massive companion.  In particular, we suggest the \ion{Si}{3} and 
other lines found in the spectral reconstruction for orbital motion
of the massive companion (\S6) are probably formed in the extended 
accretion disk ``photosphere'' surrounding the massive companion. 
These features are very broad (indicative of the large Keplerian motions 
in the disk) and appear similar to those in early B-supergiants 
(suggesting a gas density that is relatively low compared to most stellar 
photospheres).  Furthermore, the fact that this spectral component 
appears slightly cooler (like a B0.5~I star) than the supergiant (O9.7~Ibpe)
is consistent with idea that the gas has cooled slightly since leaving
the supergiant and is shielded from the flux from the presumably hot 
massive companion by the denser regions of the accretion disk.  
Similar kinds of disk photospheric features have recently been 
detected in the spectrum of $\beta$~Lyr \citep{ak07}.

Minimum light according to the recent ephemeris from \citet{kre04} 
occurs at spectroscopic phase $\phi = 0.980 \pm 0.008$ (Table~\ref{OrbSol}).
The fact that the deeper eclipse occurs when the supergiant is 
behind the massive component is consistent with the somewhat 
cooler disk photosphere we find from the line patterns in 
the reconstructed spectrum.  

The large mass of the massive companion suggests the star probably
has a vigorous radiatively-driven stellar wind.  Hydrodynamical models for $\beta$~Lyr 
indicate the radiatively-driven wind of the mass gainer will break out of the accretion 
torus in bipolar flows that are the suspected source of the H$\alpha$ 
emission in $\beta$~Lyr \citep{har96,ak07}.  If bipolar wind outflows are 
also the source of some of the H$\alpha$ emission in RY~Scuti, then they should extend
far away from the orbital plane and be minimally eclipsed at conjunction
phases.  We may test this idea by comparing the orbital phase variations in
the H$\alpha$ equivalent width ($W_\lambda$) with the $V$-band light curve. 
We measure the total emission flux relative to a time variable continuum 
flux and if the emission flux is constant, then the product of the emission
equivalent width and continuum flux will also be constant
(i.e., the measured equivalent width will vary inversely as the continuum flux). 
In Figure~\ref{lightcurve}, we show the orbital phase variations in $W_\lambda$(H$\alpha$)
normalized to their mean at the quadrature phases.  We see the measured 
$W_\lambda$ increases during the decreases in flux that occur surrounding the
times of eclipse at $\phi=0.0$ and 0.5.  We also show in Figure~\ref{lightcurve} 
the inverse continuum flux variation derived from the $V$-band observations of 
\citet{djur01} and transformed by the equation described in \S3.
The fact the broad H$\alpha$ equivalent width variations follow the inverse 
light curve so closely indicates the H$\alpha$ emission source is not 
substantially eclipsed by either the supergiant or the accretion disk, 
and therefore probably forms over a volume that is either very large 
and/or extended far from the orbital plane (we discuss this further below).
We showed in \S5 that the bisector velocities
of the H$\alpha$ emission wings (corresponding to the fastest moving gas)
show approximately the same velocity pattern as expected for the massive 
companion and this anti-phase velocity pattern is seen in Figure~\ref{HaCoude}.
The combined evidence of orbital motion
and the lack of eclipses suggests the emission line wings of
the H$\alpha$ feature (and probably part of the main core) form
in the bipolar outflows from the disk-constrained wind of the
massive companion.  The highest projected speeds observed ($\approx
500$ km~s$^{-1}$) are consistent with outflow velocities normal
to the orbital plane of $\approx 2000$ km~s$^{-1}$, a typical value
among O-dwarf winds.

The central part of the H$\alpha$ profile appears as a quasi-P~Cygni
profile (with an emission peak at redshifted velocities and a weak
absorption at blueshifted velocities) that usually implies a gas outflow.
This part of the H$\alpha$ profile is generally stationary, showing 
neither the motion associated with the supergiant nor that of the 
massive companion.  This is very different from the case of the wind emission
of the comparable supergiant in the massive X-ray
binary system Cygnus~X-1, where the H$\alpha$
emission is much weaker and follows the Doppler motions of the supergiant
\citep{gie03}.  Furthermore, the wind emission in single
OB-supergiants is generally an order of magnitude weaker than observed
in RY~Scuti \citep{mor04}, which indicates that the dense outflow
in RY~Scuti occurs over a volume much larger than that in the region
immediately surrounding the supergiant.  
The lack of any orbital velocity variation suggests 
either the source is close to the center of mass or the flux originates 
far from the center of mass where any orbital motions disappear due to 
conservation of angular momentum.
The first possibility appears to be ruled out by the fact that the main 
core shows no evidence of eclipses (Fig.~\ref{lightcurve}).  Taken together, 
these properties suggest that a large fraction of the emission flux 
originates in outflowing gas that surrounds the binary as a whole.  This
circumbinary disk is probably fed by the gas flows leaving the binary in
the vicinity of the L2 and L3 regions.  These broad flows may have
sufficient vertical extent to account for the blueshifted absorption
observed at the quadrature phases.  The absence of any substantial
decrease in emission flux around $\phi=0.0$ (Fig.~\ref{lightcurve}) 
suggests that the inner radius of the circumbinary disk is larger than
$a_{SG}+R_{SG}/\cos i = 131 R_\odot$, i.e., approximately twice
the binary separation.  On the other hand, since the gas density
will decline with increasing radius and since the H$\alpha$ emission
depends on gas density squared, we expect that most of the emission comes
from the inner part of the circumbinary disk.  Thus, the H$\alpha$
emitting region of the circumbinary disk probably has a characteristic
radius of approximately 1~AU.  The circumbinary disk flux contribution
appears to decline as expected among the profiles of the higher members 
of the Balmer sequence (see Fig.~16 in \citealt{smi02}).  
We note for completeness that the broad asymmetrical shape of the 
H$\alpha$ emission from the inner circumbinary disk appears somewhat 
like a velocity-expanded version of the profile from the outer 
double-toroidal nebula \citep{smi02}.  Since the latter is due to a 
spatial gas asymmetry in the outer expanding ring, it is possible 
that there may also be asymmetries in the mass loss into the 
inner circumbinary disk.

\placefigure{lightcurve}  

The \ion{He}{2} $\lambda 4686$ emission found in the spectrum of RY~Scuti 
has a radial velocity curve that is similar to that of the massive companion 
(Fig.~\ref{RV-all}) and this suggests it is formed in the same vicinity.  
Furthermore, the equivalent width variations of the 
\ion{He}{2} $\lambda 4686$ emission (Fig.~\ref{lightcurve})
indicate a drop in relative emission strength at $\phi=0.5$ when the 
supergiant is in the foreground, consistent with an origin near the massive
companion.  The hydrodynamical models \citep{na06b} suggest
that the hottest gas is located at the place where some of the incoming 
gas stream through L1 climbs over the accretion torus vertical extensions and 
collides with the high speed wind from the massive companion.  Thus, the hot 
gas is predicted to have a velocity curve for a position slightly offset 
from the centroid of the massive companion towards the binary center of mass.
This appears to be consistent with the \ion{He}{2} $\lambda 4686$ velocity
curve (Fig.~\ref{RV-all}) that has a slightly smaller semiamplitude 
than that for the \ion{Si}{3} lines that probably form in the disk photosphere
(which is approximately centered on the massive companion).  
 
Our results indicate the mass outflow from RY~Scuti has at least two 
components, a bipolar outflow of hot stellar wind gas from the 
massive companion and an equatorial outflow of cooler gas fed by 
the plume from the L2 region and disk leakage from the L3 region.
There is no direct evidence of the high-speed, bipolar outflow 
in H$\alpha$ images made with {\it HST} (shown in \citealt{smi99}), but 
future optical and radio emission maps made with much higher 
angular resolution may reveal these jet-like features. 
We suspect most of the cool gas outflow in the orbital 
plane ends up in a circumbinary disk that slowly transports 
mass and angular momentum away from the binary and may eventually be a
source of gas for the outer double-toriodal nebula.  This draining
of angular momentum from the binary may explain the 
observed shrinking of the orbital period and semimajor axis 
\citep{sim99}.  It is possible 
the density and spatial extent of the circumbinary disk 
is sufficient to become a significant opacity source blocking 
the ionizing flux from the central binary.  We speculate that 
this may be one of the reasons why the surrounding H~II double-toroidal
nebula seen in {\it HST} images has a darker equatorial zone 
(shaded from ionizing radiation by the intervening circumbinary disk).


\section{Conclusions}

Our spectroscopic investigation of RY~Scuti has led to a reassessment 
of the component masses, a new understanding of the properties of 
the spectral features related to each star, and a sketch of how 
the binary is ejecting gas from the system.  Our new radial velocities
and orbital elements for the supergiant improve upon the past work
and we argue that the orbit is probably circular as expected for 
Roche-filling systems.  We determined radial velocities for the 
\ion{Si}{3} $\lambda 4552$ absorption line, a feature that is probably 
formed in the vicinity of massive companion, and by combining the 
semiamplitudes of the supergiant and this feature, we estimate the mass 
ratio is $M_{MC}/M_{SG} = 4.2 \pm 0.7$.  
If we adopt an orbital inclination from that for the surrounding 
double-toroidal nebula, then the stellar masses are  
$M_{SG}=7.1 \pm 1.2 M_\odot$ and $M_{MC}=30.0 \pm 2.1 M_\odot$.
These are consistent with the idea that the massive companion is an O-star 
hidden within a dense accretion disk and that the mass-losing supergiant
is destined to become a Wolf-Rayet star.  In the distant future, this system 
may resemble the well known O + WR binary $\gamma^2$~Vel that has component masses
of $M(O) = 28.5 \pm 1.1 M_\odot$ and $M(WR) = 9.0\pm0.6$ \citep{nor07}.

We used a Doppler tomographic reconstruction scheme to separate the 
component optical spectra of the stars.  The line patterns of the 
the supergiant dominate the combined spectrum and a comparison with 
a similarly classified star indicates that the supergiant contributes
about $57\%$ of the visible flux.  The optical spectrum of the supergiant 
displays unusually weak carbon lines, suggesting
the photosphere (like the surrounding nebula) contains CNO-processed gas. 
The reconstructed spectrum for the massive companion has a superficial 
resemblance to that of a rapidly rotating B-supergiant, but we argue this 
spectrum probably forms in the accretion disk photosphere surrounding the
massive companion. 

We found several lines of evidence indicating mass loss from the binary. 
The apparent strengthening of blueshifted parts of many absorption lines at 
both conjunctions and especially following supergiant inferior conjunction 
suggest that the supergiant is losing mass in broad streams directed 
both towards and away from the massive companion.  The strong H$\alpha$ 
feature appears as broad P~Cygni line indicating outflow.  The radial velocity 
curve of the H$\alpha$ wings indicates the highest velocity emission originates
near the massive companion, probably in a bipolar outflow where the stellar 
wind of the massive companion breaks out of the thick surrounding accretion disk.
The core H$\alpha$ emission is stationary and uneclipsed, and we suggest it 
forms mainly in an outflowing equatorial region (a circumbinary disk).  
The other important emission line, \ion{He}{2} $\lambda 4686$, 
also has a radial velocity curve consistent with an origin near the massive 
companion, and we suggest it forms in hot regions where gas from the L1 stream 
flows over the thicker portions of the disk and strikes the fast wind from 
the massive companion. 

We argue that the gas lost from the supergiant through the outer L2 region 
and also lost from the outer parts of the accretion disk near L3 ends up in a 
relatively cool and dense circumbinary disk with a characteristic H$\alpha$
emitting radius of 1 AU.  We suspect this disk gas  
will continue to flow outwards and may help channel gas into the surrounding 
double-toroidal nebula.  \citet{smi01} studied the proper motions of 
the outer double-torus nebula and they conclude the gas in this H~II 
region was probably ejected from the central binary around the year $1876 \pm 20$.
It is possible this time corresponded to an epoch with an especially high binary 
mass transfer rate and associated outflow into the circumbinary disk. 
We note, for example, that the first light curve for the binary 
from \citet{gap37} (covering the years 1924 to 1934) shows a secondary minimum 
that is only 0.3~mag deep compared to the current day observed 0.5~mag drop
\citep{djur01} while the primary eclipse depth is the same.  This suggests 
the disk surrounding the massive companion was cooler in the early 
part of the last century, which would be consistent with the presence of 
a thicker, more extended disk resulting from a higher mass transfer rate. 
Finally, we note that the mass loss into the circumbinary disk removes 
angular momentum from the central binary, consistent with the observation
that the orbital period is decreasing \citep{sim99}.  Thus, contrary to 
the predictions of conservative mass transfer models, the components of RY~Scuti
are continuing to be drawn together even after the mass ratio reversal has occurred. 
We may therefore expect that an episode of more intense interaction and 
mass transfer and mass loss lies in the immediate future for RY~Scuti, 
and continuing observations will be vital in documenting this key stage in 
the evolution of massive binaries.


\acknowledgments

We thank Daryl Willmarth and the staffs of KPNO and CTIO for 
their support in making these observations possible.  
This material is based upon work supported by the National Science Foundation
under Grants No.~AST-0506573 and AST-0606861.
Institutional support has been provided from the GSU College
of Arts and Sciences and from the Research Program Enhancement
fund of the Board of Regents of the University System of Georgia,
administered through the GSU Office of the Vice President for Research.
RDG is supported by NASA, the NSF, and the US Air Force.
We gratefully acknowledge all this support.



\clearpage


\begin{deluxetable}{cccccl}
\tabletypesize{\scriptsize}
\tablewidth{0pc}
\tablecaption{Journal of Spectroscopy \label{ScopeInfo}}
\tablehead{
\colhead{Run} &
\colhead{Dates} &
\colhead{Range} &
\colhead{Resolving Power} &
\colhead{Number of} &
\colhead{Observatory/Telescope/} \\
\colhead{Number} &
\colhead{(HJD-2,450,000)} &
\colhead{(\AA)} &
\colhead{($\lambda/\triangle\lambda$)} &
\colhead{Spectra} &
\colhead{Spectral Grating/CCD}}
\startdata
1 \dotfill &1354.7 -- 1354.8   &   6431 -- 6785    &\phn5440    & \phn2  & KPNO/0.9m/RC181/TI5  \\
2 \dotfill &1355.7 -- 1364.9   &   5405 -- 6743    &\phn3950    &    32  & KPNO/0.9m/RC181/F3KB \\
3 \dotfill &1373.7 -- 1394.6   &   3600 -- 9200    &   48000    &    17  & ESO~/1.5m/FEROS/EEV $2\times4$K \\
4 \dotfill &1421.8 -- 1429.7   &   5397 -- 6735    &\phn4050    & \phn6  & KPNO/0.9m/RC181/F3KB \\
5 \dotfill &3152.9 -- 3164.9   &   4068 -- 4738    &\phn2430    &    10  & CTIO/1.5m/\#47II/Loral 1K \\
\enddata
\end{deluxetable}

\begin{deluxetable}{cl}
\tablewidth{0pt}
\tablenum{2}
\tablecaption{Supergiant Radial Velocity Line Sample\label{linelist}}
\tablehead{
\colhead{Telescope Run} &
\colhead{Line Regions}}
\startdata
ESO 1.5 m  &  \ion{Si}{4} $\lambda4088$; \ion{Si}{4} $\lambda4116$, \ion{He}{1} $\lambda4121$; \\
           &  \ion{N}{3} $\lambda\lambda4510,4514,4518$; \\
CTIO 1.5 m &  \ion{Si}{4} $\lambda4088$; \ion{Si}{4} $\lambda4116$, \ion{He}{1} $\lambda4121$; \\
           &  \ion{N}{3} $\lambda\lambda4510,4514,4518$; \\
KPNO CF    &  \ion{N}{2} $\lambda6610$ (em.); \ion{Si}{4} $\lambda6701$ (em.) \\
\enddata
\end{deluxetable}

\begin{deluxetable}{lccccl}
\tabletypesize{\scriptsize}
\tablewidth{0pt}
\tablenum{3}
\tablecaption{Supergiant Radial Velocity Measurements\label{SGrv}}
\tablehead{
\colhead{HJD}            &
\colhead{Orbital}        &
\colhead{$V_{r}$}	 &
\colhead{$\sigma$}       &
\colhead{$O-C$}          &
\colhead{}               \\
\colhead{($-$2,450,000)} &
\colhead{Phase}          &
\colhead{(km~s$^{-1}$)}  &
\colhead{(km~s$^{-1}$)}  & 
\colhead{(km~s$^{-1}$)}  & 
\colhead{Telescope}      }
\startdata 
  1354.781 \dotfill &  0.231 & 
             $-221.6$ &       11.6 & \phs     $  14.6$ & KPNO CF \\
  1354.803 \dotfill &  0.233 & 
             $-218.0$ &\phn    9.6 & \phs     $  18.5$ & KPNO CF \\
  1355.674 \dotfill &  0.312 & 
             $-217.2$ &       30.0 & \phn\phs $   2.2$ & KPNO CF \\
  1355.701 \dotfill &  0.314 & 
             $-207.2$ &       11.3 & \phs     $  10.8$ & KPNO CF \\
  1355.914 \dotfill &  0.333 & 
             $-215.4$ &       16.9 &          $ -10.8$ & KPNO CF \\
  1355.935 \dotfill &  0.335 & 
             $-212.4$ &       27.4 & \phn     $  -9.3$ & KPNO CF \\
  1356.722 \dotfill &  0.406 & 
             $-116.8$ &       20.1 & \phs     $  10.9$ & KPNO CF \\
  1356.744 \dotfill &  0.408 & 
             $-122.4$ &       25.1 & \phn\phs $   2.6$ & KPNO CF \\
  1356.910 \dotfill &  0.423 & 
             $-122.0$ &       28.7 &          $ -17.0$ & KPNO CF \\
  1357.691 \dotfill &  0.493 & 
\phn         $ -30.1$ &       12.2 &          $ -30.5$ & KPNO CF \\
  1357.712 \dotfill &  0.495 & 
\phn\phn     $  -7.8$ &       10.1 &          $ -11.1$ & KPNO CF \\
  1357.832 \dotfill &  0.506 & 
\phn\phn\phs $   8.5$ &       13.4 &          $ -11.7$ & KPNO CF \\
  1357.853 \dotfill &  0.507 & 
\phn\phs     $  25.2$ &       34.2 & \phn\phs $   2.0$ & KPNO CF \\
  1357.912 \dotfill &  0.513 & 
\phn\phs     $  21.5$ &\phn    9.7 & \phn     $  -9.9$ & KPNO CF \\
  1357.933 \dotfill &  0.515 & 
\phn\phs     $  29.7$ &\phn    9.6 & \phn     $  -4.7$ & KPNO CF \\
  1358.861 \dotfill &  0.598 & 
\phs         $ 127.5$ &\phn    9.6 &          $ -28.1$ & KPNO CF \\
  1359.663 \dotfill &  0.670 & 
\phs         $ 232.0$ &       26.9 & \phn\phs $   2.0$ & KPNO CF \\
  1359.709 \dotfill &  0.674 & 
\phs         $ 214.8$ &\phn    9.8 &          $ -18.3$ & KPNO CF \\
  1359.889 \dotfill &  0.691 & 
\phs         $ 213.2$ &       13.3 &          $ -30.4$ & KPNO CF \\
  1359.911 \dotfill &  0.692 & 
\phs         $ 223.3$ &       15.7 &          $ -21.4$ & KPNO CF \\
  1360.664 \dotfill &  0.760 & 
\phs         $ 251.5$ &       10.7 & \phn     $  -8.8$ & KPNO CF \\
  1360.730 \dotfill &  0.766 & 
\phs         $ 248.9$ &       17.6 &          $ -10.6$ & KPNO CF \\
  1360.860 \dotfill &  0.778 & 
\phs         $ 238.7$ &       25.9 &          $ -18.3$ & KPNO CF \\
  1360.881 \dotfill &  0.780 & 
\phs         $ 242.1$ &       27.4 &          $ -14.4$ & KPNO CF \\
  1361.872 \dotfill &  0.869 & 
\phs         $ 209.2$ &       20.6 & \phs     $  14.7$ & KPNO CF \\
  1361.953 \dotfill &  0.876 & 
\phs         $ 199.8$ &\phn   10.0 & \phs     $  13.2$ & KPNO CF \\
  1362.661 \dotfill &  0.940 & 
\phs         $ 142.8$ &\phn    9.6 & \phs     $  39.1$ & KPNO CF \\
  1362.706 \dotfill &  0.944 & 
\phs         $ 120.8$ &       15.2 & \phs     $  22.9$ & KPNO CF \\
  1363.684 \dotfill &  0.032 & 
\phn\phn\phs $   0.8$ &\phn    9.6 & \phs     $  38.5$ & KPNO CF \\
  1363.807 \dotfill &  0.043 & 
\phn         $ -43.5$ &\phn    9.6 & \phs     $  11.0$ & KPNO CF \\
  1363.839 \dotfill &  0.046 & 
\phn         $ -61.5$ &\phn    9.6 & \phn     $  -2.6$ & KPNO CF \\
  1363.925 \dotfill &  0.053 & 
\phn         $ -70.6$ &\phn    9.6 & \phn     $  -0.1$ & KPNO CF \\
  1363.946 \dotfill &  0.055 & 
\phn         $ -75.4$ &\phn    9.6 & \phn     $  -2.1$ & KPNO CF \\
  1364.900 \dotfill &  0.141 & 
             $-191.6$ &       31.9 &          $ -10.1$ & KPNO CF \\
  1373.693 \dotfill &  0.931 & 
\phs         $ 112.4$ &       12.0 & \phn     $  -3.3$ & ESO 1.5 m \\
  1373.758 \dotfill &  0.937 & 
\phs         $ 103.7$ &\phn    8.6 & \phn     $  -3.7$ & ESO 1.5 m \\
  1374.655 \dotfill &  0.018 & 
\phn         $ -53.7$ &\phn    7.3 &          $ -37.3$ & ESO 1.5 m \\
  1375.709 \dotfill &  0.113 & 
             $-161.2$ &\phn    6.5 &          $ -10.6$ & ESO 1.5 m \\
  1379.718 \dotfill &  0.473 & 
\phn         $ -16.1$ &\phn    7.4 & \phs     $  14.6$ & ESO 1.5 m \\
  1381.646 \dotfill &  0.646 & 
\phs         $ 214.1$ &       11.5 & \phn\phs $   4.4$ & ESO 1.5 m \\
  1382.668 \dotfill &  0.738 & 
\phs         $ 272.0$ &\phn    3.9 & \phs     $  11.9$ & ESO 1.5 m \\
  1383.645 \dotfill &  0.826 & 
\phs         $ 252.3$ &\phn    8.5 & \phs     $  19.4$ & ESO 1.5 m \\
  1384.625 \dotfill &  0.914 & 
\phs         $ 140.1$ &\phn    6.8 & \phn\phs $   0.5$ & ESO 1.5 m \\
  1385.687 \dotfill &  0.010 & 
\phn         $ -53.4$ &       10.9 &          $ -49.9$ & ESO 1.5 m \\
  1386.650 \dotfill &  0.096 & 
             $-161.5$ &\phn    6.9 &          $ -31.5$ & ESO 1.5 m \\
  1388.583 \dotfill &  0.270 & 
             $-215.0$ &\phn    3.4 & \phs     $  20.9$ & ESO 1.5 m \\
  1390.664 \dotfill &  0.457 & 
\phn         $ -49.4$ &\phn    8.9 & \phn\phs $   5.8$ & ESO 1.5 m \\
  1391.602 \dotfill &  0.541 & 
\phn\phs     $  78.5$ &       12.8 & \phn\phs $   3.2$ & ESO 1.5 m \\
  1392.619 \dotfill &  0.633 & 
\phs         $ 194.5$ &       10.4 & \phn     $  -1.6$ & ESO 1.5 m \\
  1393.638 \dotfill &  0.724 & 
\phs         $ 272.4$ &\phn    7.1 & \phs     $  14.9$ & ESO 1.5 m \\
  1394.615 \dotfill &  0.812 & 
\phs         $ 253.2$ &\phn    1.4 & \phs     $  11.2$ & ESO 1.5 m \\
  1421.765 \dotfill &  0.253 & 
             $-230.9$ &\phn    9.6 & \phn\phs $   6.9$ & KPNO CF \\
  1425.784 \dotfill &  0.614 & 
\phs         $ 165.2$ &\phn    9.6 & \phn     $  -9.9$ & KPNO CF \\
  1426.745 \dotfill &  0.700 & 
\phs         $ 236.6$ &       29.3 &          $ -12.1$ & KPNO CF \\
  1427.748 \dotfill &  0.790 & 
\phs         $ 244.5$ &       27.8 & \phn     $  -8.2$ & KPNO CF \\
  1428.718 \dotfill &  0.878 & 
\phs         $ 179.7$ &       24.1 & \phn     $  -5.0$ & KPNO CF \\
  1429.713 \dotfill &  0.967 & 
\phn\phs     $  98.4$ &       21.3 & \phs     $  35.8$ & KPNO CF \\
  3152.930 \dotfill &  0.871 & 
\phs         $ 195.5$ &       13.9 & \phn\phs $   3.0$ & CTIO 1.5 m \\
  3153.825 \dotfill &  0.951 & 
\phn\phs     $  83.9$ &       22.5 & \phn     $  -3.0$ & CTIO 1.5 m \\
  3154.927 \dotfill &  0.050 & 
             $-118.3$ &       46.0 &          $ -52.4$ & CTIO 1.5 m \\
  3156.922 \dotfill &  0.229 & 
             $-216.3$ &       15.4 & \phs     $  19.5$ & CTIO 1.5 m \\
  3157.932 \dotfill &  0.320 & 
             $-204.4$ &       17.0 & \phn\phs $   9.6$ & CTIO 1.5 m \\
  3158.928 \dotfill &  0.410 & 
             $-128.2$ &\phn    2.5 & \phn     $  -5.9$ & CTIO 1.5 m \\
  3161.915 \dotfill &  0.678 & 
\phs         $ 239.7$ &       14.6 & \phn\phs $   3.8$ & CTIO 1.5 m \\
  3162.914 \dotfill &  0.768 & 
\phs         $ 254.1$ &       11.2 & \phn     $  -5.0$ & CTIO 1.5 m \\
  3163.892 \dotfill &  0.856 & 
\phs         $ 206.1$ &       17.7 & \phn     $  -1.4$ & CTIO 1.5 m \\
  3164.903 \dotfill &  0.947 & 
\phn\phs     $  83.5$ &       19.6 & \phn     $  -9.5$ & CTIO 1.5 m \\
\enddata
\end{deluxetable}

\begin{deluxetable}{lcccc}
\tabletypesize{\scriptsize}
\tablewidth{0pc}
\tablenum{4}
\tablecaption{Orbital Elements for RY Scuti\label{OrbSol}}
\tablehead{
\colhead{Element}               & 
\colhead{Circular}              & 
\colhead{Elliptical}            & 
\colhead{Sahade et al.\ (2002)}	&
\colhead{Skul'skii (1992)}	}
\startdata
$P$~(days)                  \dotfill & 11.12445\tablenotemark{a}  & 11.12445\tablenotemark{a}  & 11.124646\tablenotemark{a}  & 11.1250\tablenotemark{a} \\
$T_{SC}$ (HJD--2,400,000)   \dotfill & $51396.71 \pm 0.02$  & \nodata     & \nodata          & $44777.9 \pm1.0$\\
$T$ (HJD--2,400,000)        \dotfill & \nodata         & $51395.68 \pm 0.40$ & 45107.74      & \nodata         \\
$e$                         \dotfill & 0               & $0.05 \pm 0.01$  & $0.16 \pm 0.04$  & 0               \\
$\omega$ (deg)              \dotfill & \nodata         & $57 \pm 13$      & $345 \pm 15$     & \nodata         \\
$K_{SG}$ (km~s$^{-1}$)      \dotfill & $249 \pm 3$     & $247 \pm 3$      & $273 \pm 14$     & $235 \pm 3$     \\
$K_{MC}$ (km~s$^{-1}$)      \dotfill & $ 59 \pm 9$     & \nodata          & $ 71 \pm  9$     & $ 71 \pm 2$     \\
$\gamma_{SG}$ (km~s$^{-1}$) \dotfill & $ 11 \pm 2$     & $12 \pm 2$       & $ 17 \pm  5$     & $  9 \pm 2$     \\
$\gamma_{MC}$ (km~s$^{-1}$) \dotfill & $-20 \pm 9$     & \nodata          & \nodata          & $ 11 \pm 2$     \\
r.m.s.$_{SG}$ (km~s$^{-1}$) \dotfill & 19              & 17               & \nodata          & \nodata         \\
r.m.s.$_{MC}$ (km~s$^{-1}$) \dotfill & 25              & \nodata          & \nodata          & \nodata         \\
$a$ ($R_\odot$)             \dotfill & $69.9\pm 2.3$   & \nodata          & $77  \pm 4$      & 68              \\
$q=M_{MC}/M_{SG}$           \dotfill & $4.2 \pm 0.7$   & \nodata          & $ 3.9\pm 0.5$    & $3.3 \pm 0.1$   \\
$M_{SG}$ ($M_\odot$)        \dotfill & $7.1 \pm 1.2$   & \nodata          & $10.2\pm 1.7$    &  8              \\
$M_{MC}$ ($M_\odot$)        \dotfill & $30.0\pm 2.1$   & \nodata          & $39  \pm 4$      & 26              \\
\enddata
\tablenotetext{a}{Fixed}
\end{deluxetable}

\begin{deluxetable}{lccc}
\tablewidth{0pt}
\tabletypesize{\scriptsize}
\tablenum{5}
\tablecaption{H$\alpha$ Wing Velocities and Equivalent Widths\label{Hawings}}
\tablehead{
\colhead{Date}                 &
\colhead{Orbital}              &
\colhead{$V_r$}                &
\colhead{$W_\lambda$}          \\
\colhead{(HJD$-$2,450,000)}    &
\colhead{Phase}                &
\colhead{(km~s$^{-1}$)}        &
\colhead{(\AA )}               }
\startdata
 1354.781\dotfill  &  0.231  & \phn  83.3  &  $-$13.4  \\
 1354.803\dotfill  &  0.233  & \phn  82.9  &  $-$13.4  \\
 1355.674\dotfill  &  0.312  & \phn  76.8  &  $-$12.9  \\
 1355.701\dotfill  &  0.314  & \phn  73.4  &  $-$13.2  \\
 1355.913\dotfill  &  0.333  & \phn  74.2  &  $-$13.2  \\
 1355.935\dotfill  &  0.335  & \phn  72.3  &  $-$12.7  \\
 1356.722\dotfill  &  0.406  & \phn  60.3  &  $-$14.8  \\
 1356.744\dotfill  &  0.408  & \phn  61.0  &  $-$14.8  \\
 1356.910\dotfill  &  0.423  & \phn  59.0  &  $-$14.6  \\
 1357.691\dotfill  &  0.493  & \phn  42.6  &  $-$21.9  \\
 1357.712\dotfill  &  0.495  & \phn  44.1  &  $-$22.8  \\
 1357.832\dotfill  &  0.506  & \phn  38.8  &  $-$23.0  \\
 1357.853\dotfill  &  0.507  & \phn  38.0  &  $-$23.0  \\
 1357.912\dotfill  &  0.513  & \phn  35.7  &  $-$20.5  \\
 1357.933\dotfill  &  0.515  & \phn  35.4  &  $-$19.6  \\
 1358.861\dotfill  &  0.598  & \phn  30.6  &  $-$14.5  \\
 1359.663\dotfill  &  0.670  & \phn  20.8  &  $-$12.5  \\
 1359.709\dotfill  &  0.674  & \phn  23.7  &  $-$13.4  \\
 1359.889\dotfill  &  0.691  & \phn  21.9  &  $-$12.1  \\
 1359.911\dotfill  &  0.692  & \phn  21.8  &  $-$12.3  \\
 1360.664\dotfill  &  0.760  & \phn  26.4  &  $-$13.6  \\
 1360.730\dotfill  &  0.766  & \phn  26.3  &  $-$14.0  \\
 1360.860\dotfill  &  0.778  & \phn  33.4  &  $-$13.0  \\
 1360.881\dotfill  &  0.780  & \phn  33.8  &  $-$12.7  \\
 1361.872\dotfill  &  0.869  & \phn  38.5  &  $-$13.8  \\
 1361.953\dotfill  &  0.876  & \phn  40.2  &  $-$12.1  \\
 1362.661\dotfill  &  0.940  & \phn  27.3  &  $-$18.8  \\
 1362.706\dotfill  &  0.944  & \phn  25.7  &  $-$19.5  \\
 1363.684\dotfill  &  0.032  & \phn  49.5  &  $-$21.9  \\
 1363.807\dotfill  &  0.043  & \phn  55.2  &  $-$21.3  \\
 1363.839\dotfill  &  0.046  & \phn  56.9  &  $-$21.1  \\
 1363.925\dotfill  &  0.053  & \phn  66.0  &  $-$18.6  \\
 1363.946\dotfill  &  0.055  & \phn  70.3  &  $-$18.0  \\
 1364.900\dotfill  &  0.141  &      119.1  &  $-$13.6  \\
 1373.693\dotfill  &  0.931  & \phn  24.4  &  $-$18.4  \\
 1373.758\dotfill  &  0.937  & \phn  24.4  &  $-$19.2  \\
 1374.655\dotfill  &  0.018  & \phn  39.5  &  $-$21.5  \\
 1375.709\dotfill  &  0.113  & \phn  90.6  &  $-$15.1  \\
 1379.718\dotfill  &  0.473  & \phn  41.7  &  $-$19.7  \\
 1381.646\dotfill  &  0.646  & \phn  12.7  &  $-$12.7  \\
 1382.668\dotfill  &  0.738  & \phn\phn5.5 &  $-$11.5  \\
 1383.645\dotfill  &  0.826  & \phn  14.2  &  $-$12.4  \\
 1384.625\dotfill  &  0.914  & \phn  15.2  &  $-$15.1  \\
 1385.687\dotfill  &  0.010  & \phn  34.4  &  $-$21.3  \\
 1386.650\dotfill  &  0.096  & \phn  87.7  &  $-$16.1  \\
 1388.583\dotfill  &  0.270  & \phn  54.8  &  $-$11.8  \\
 1390.664\dotfill  &  0.457  & \phn  40.5  &  $-$19.0  \\
 1391.602\dotfill  &  0.541  & \phn  11.3  &  $-$18.7  \\
 1392.619\dotfill  &  0.633  & \phn  10.8  &  $-$13.2  \\
 1393.638\dotfill  &  0.724  & \phn  11.7  &  $-$13.1  \\
 1394.615\dotfill  &  0.812  & \phn  17.2  &  $-$13.0  \\
 1421.765\dotfill  &  0.253  & \phn  71.7  &  $-$12.1  \\
 1425.784\dotfill  &  0.614  & \phn  19.5  &  $-$14.3  \\
 1426.745\dotfill  &  0.700  & \phn  22.9  &  $-$12.4  \\
 1427.748\dotfill  &  0.790  & \phn  41.3  &  $-$12.4  \\
 1428.718\dotfill  &  0.878  & \phn  36.7  &  $-$13.1  \\
 1429.713\dotfill  &  0.967  & \phn  37.7  &  $-$20.9  \\
\enddata
\end{deluxetable}

\begin{deluxetable}{lcccc}
\tablewidth{0pt}
\tabletypesize{\scriptsize}
\tablenum{6}
\tablecaption{Features Associated with the Massive Companion\label{MCfeatures}}
\tablehead{
\colhead{Date}                 &
\colhead{Orbital}              &
\colhead{$V_r$(Si III)}        &
\colhead{$V_r$(He II)}         &
\colhead{$W_\lambda$(He II)}   \\
\colhead{(HJD$-$2,450,000)}    &
\colhead{Phase}                &
\colhead{(km~s$^{-1}$)}        &
\colhead{(km~s$^{-1}$)}        &
\colhead{(\AA )}               }
\startdata
 1373.693\dotfill  &  0.931  & \nodata  &  \phn 51.3  &$-$0.93  \\
 1373.758\dotfill  &  0.937  & \nodata  &  \phn 51.1  &$-$0.92  \\
 1374.655\dotfill  &  0.018  & \nodata  &  \phn 67.2  &$-$0.99  \\
 1375.709\dotfill  &  0.113  & \nodata  &  \phn 63.3  &$-$0.95  \\
 1379.718\dotfill  &  0.473  & \nodata  &  \phn 80.2  &$-$0.63  \\
 1381.646\dotfill  &  0.646  & \nodata  &  \phn 51.1  &$-$0.77  \\
 1382.668\dotfill  &  0.738  & $-$97.8  &  \phn 10.3  &$-$0.63  \\
 1383.645\dotfill  &  0.826  & $-$59.7  &  \phn 39.0  &$-$0.74  \\
 1384.625\dotfill  &  0.914  & \nodata  &  \phn 71.8  &$-$0.74  \\
 1385.687\dotfill  &  0.010  & \nodata  &  \phn 51.6  &$-$0.95  \\
 1386.650\dotfill  &  0.096  & \nodata  &  \phn 51.4  &$-$0.96  \\
 1388.583\dotfill  &  0.270  &\phs\phn3.6&     122.0  &$-$0.66  \\
 1390.664\dotfill  &  0.457  & \nodata  &  \phn 80.8  &$-$0.60  \\
 1391.602\dotfill  &  0.541  & \nodata  &  \phn\phn7.8&$-$0.74  \\
 1392.619\dotfill  &  0.633  & \nodata  &  \phn 24.7  &$-$0.81  \\
 1393.638\dotfill  &  0.724  & $-$84.6  &  \phn 15.6  &$-$0.78  \\
 1394.615\dotfill  &  0.812  & $-$59.7  &  \phn\phn1.1&$-$0.58  \\
 3152.930\dotfill  &  0.871  & \nodata  &  \phn 86.8  &$-$0.59  \\
 3153.825\dotfill  &  0.951  & \nodata  &  \phn 59.2  &$-$0.70  \\
 3154.928\dotfill  &  0.050  & \nodata  &  \phn 72.9  &$-$0.71  \\
 3156.922\dotfill  &  0.229  & \phs33.0 &      106.5  &$-$0.29  \\
 3157.932\dotfill  &  0.320  & \phs73.9 &      123.5  &$-$0.34  \\
 3158.928\dotfill  &  0.410  & \nodata  &  \phn 72.1  &$-$0.58  \\
 3161.915\dotfill  &  0.678  & \nodata  &  \phn 60.4  &$-$0.34  \\
 3162.914\dotfill  &  0.768  & $-$82.6  &  \phn 23.2  &$-$0.51  \\
 3163.892\dotfill  &  0.856  & \nodata  &  \phn 35.3  &$-$0.52  \\
 3164.903\dotfill  &  0.947  & \nodata  &  \phn 54.0  &$-$0.66  \\
\enddata
\end{deluxetable}

\begin{deluxetable}{lccc}
\tablewidth{0pt}
\tablenum{7}
\tablecaption{Summary of Radial Velocity Fits\label{MCRVfits}}
\tablehead{
\colhead{Element}                          &
\colhead{$V_r$(\ion{Si}{3})}               &
\colhead{$V_r$(\ion{He}{2} $\lambda4686$)} &
\colhead{$V_r$(H$\alpha$ wings)}           }
\startdata
$K$ (km~s$^{-1}$)      &   $59 \pm 5$ &  $36 \pm 6$ &  $30 \pm 3$  \\
$\gamma$ (km~s$^{-1}$) &  $-20 \pm 5$ &  $64 \pm 4$ &  $49 \pm 2$  \\
r.m.s. (km~s$^{-1}$)   &   22         &  21         &  13          \\
\enddata
\end{deluxetable}

\clearpage


\begin{figure}
\epsscale{.80}
\plotone{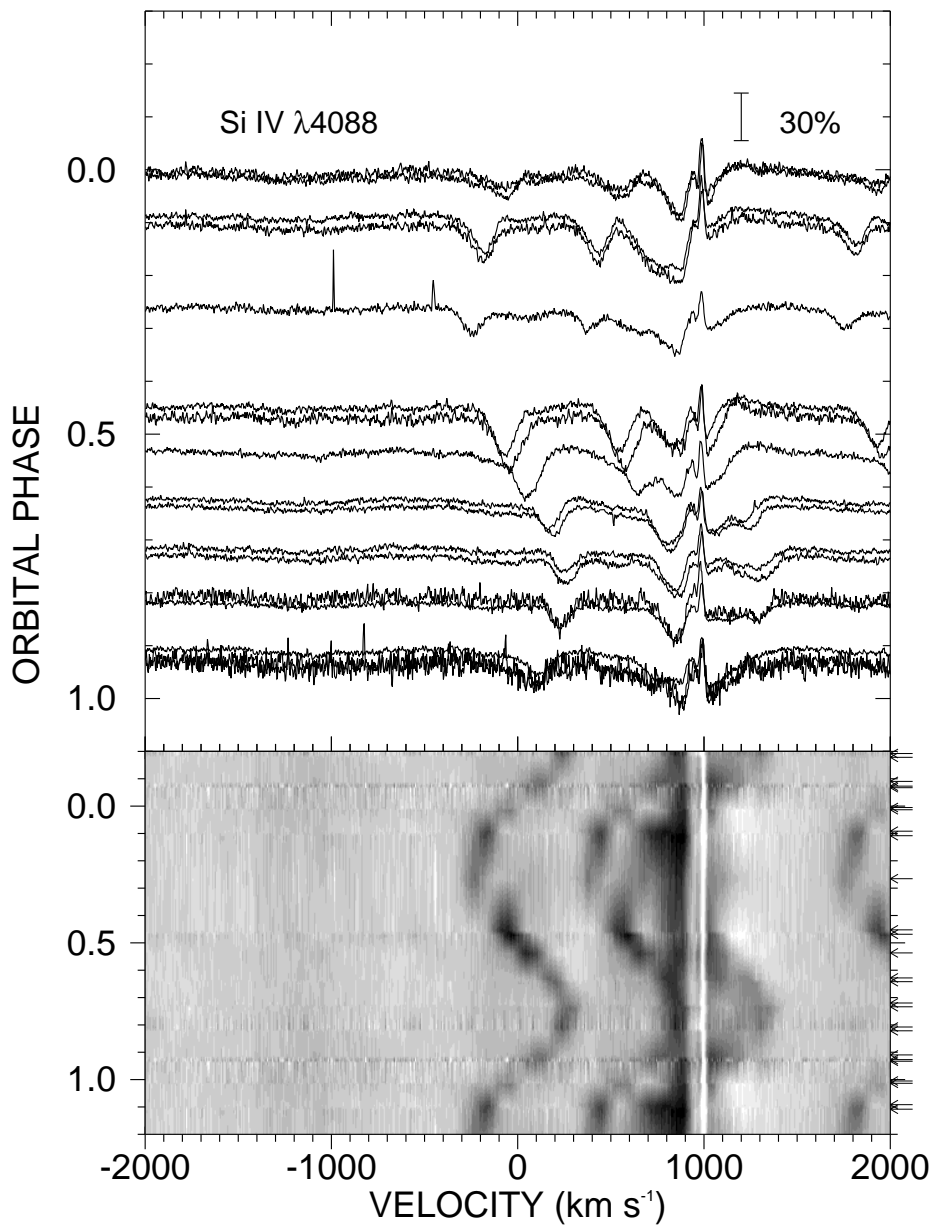}
\caption{The orbital phase variations in the \ion{Si}{4} $\lambda4088$ 
absorption line in the FEROS spectra of RY~Scuti are shown in 
linear plots ({\it top panel}) and as a gray-scale image ({\it lower panel}). 
The intensity in the gray-scale image is assigned one of 16 gray levels 
based on its value between the minimum (dark) and maximum (bright) observed values. 
The intensity between observed spectra is calculated by a linear interpolation 
between the closest observed phases (shown by arrows along the right axis).
The feature centered at 0 km~s$^{-1}$ is \ion{Si}{4} $\lambda4088$, and to
the right is \ion{N}{3} $\lambda4097$, H$\delta$ $\lambda4101$ 
(both absorption and nebular emission),
and \ion{Si}{4} $\lambda4116$ (far right edge).
\label{SiIV4088}}
\end{figure}

\begin{figure}
\epsscale{.80}
\plotone{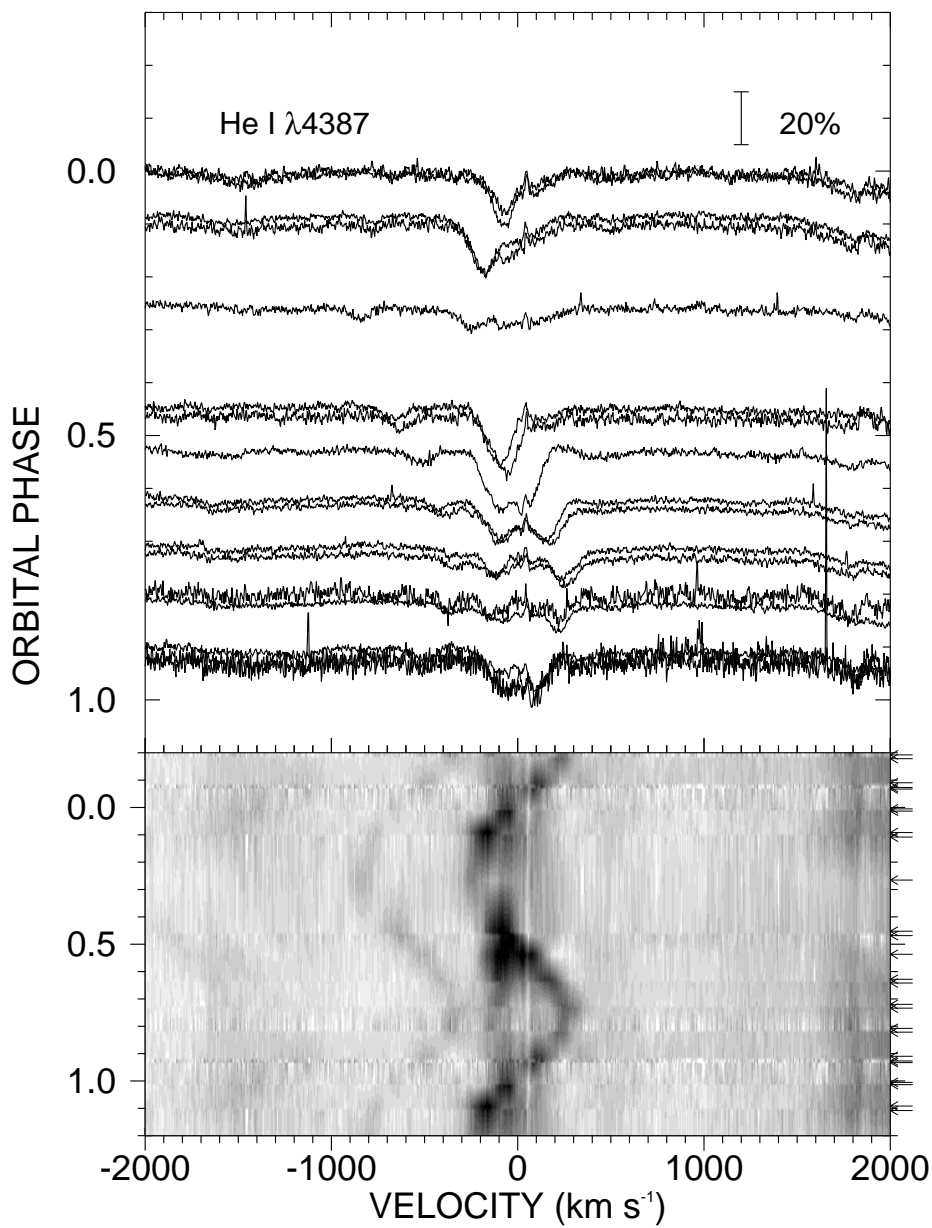}
\caption{The orbital phase variations in the \ion{He}{1} $\lambda4387$ singlet 
shown in the same format as Fig.~\ref{SiIV4088}.   This feature strengthens at conjunctions and 
develops a blueshifted absorption feature after $\phi=0.5$.
Traces of the radial velocity curve of the fainter massive companion are seen especially after
$\phi=0.5$, however, this feature is not used in the radial velocity analysis due to the 
aforementioned blueshifted absorption feature.
The fainter absorption feature just to the left is \ion{N}{3} $\lambda4379$.
\label{HeI4387}}
\end{figure}

\begin{figure}
\epsscale{.80}
\plotone{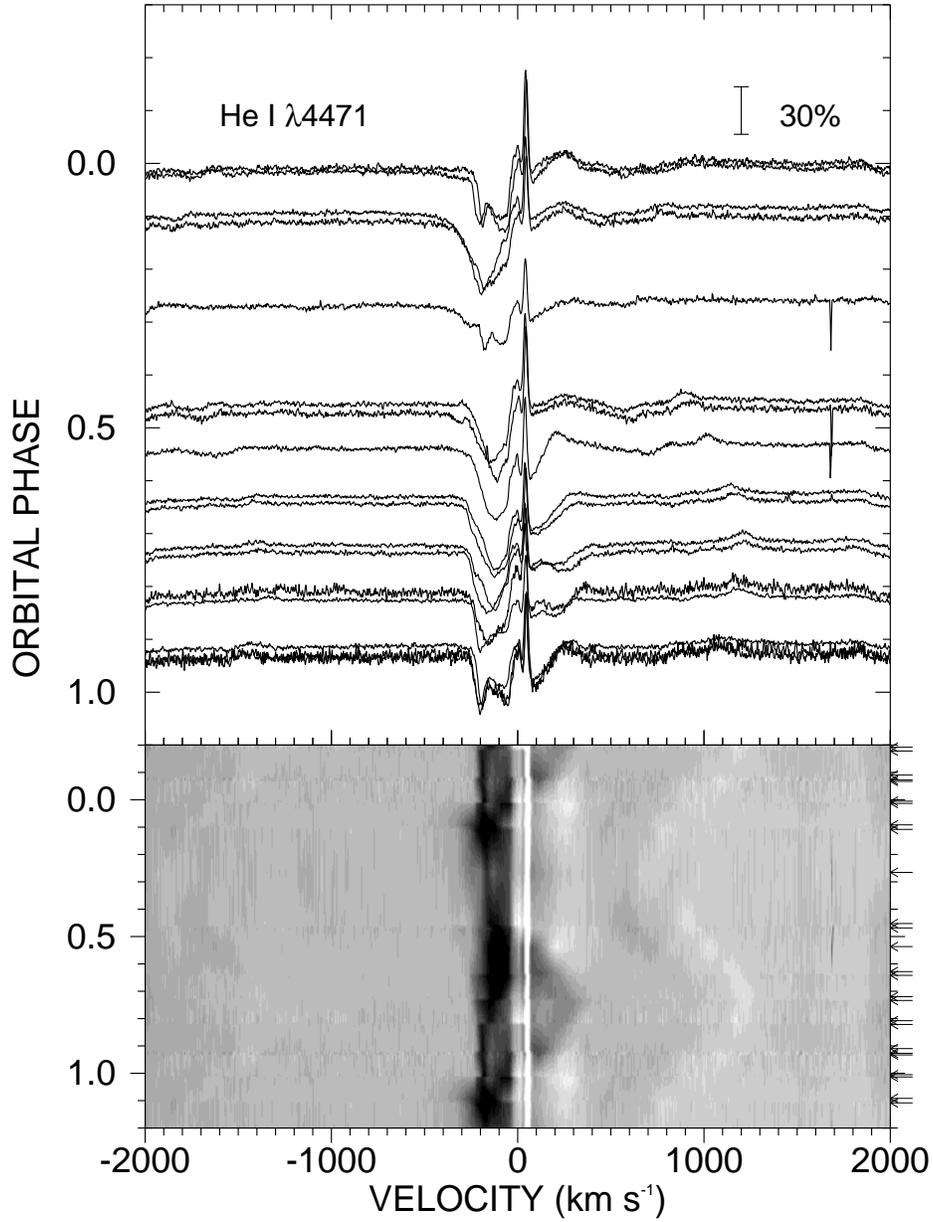}
\caption{The orbital phase variations in the \ion{He}{1} $\lambda4471$ triplet 
shown in the same format as Fig.~\ref{SiIV4088}.  This line has a stronger blueshifted absorption
feature after $\phi=0.5$ that transforms into a narrow and long-lived absorption line.
There is also some evidence of redward emission at the conjunctions that hints 
at a P~Cygni profile shape. 
\label{HeI4471}}
\end{figure}

\begin{figure}
\epsscale{.80}
\plotone{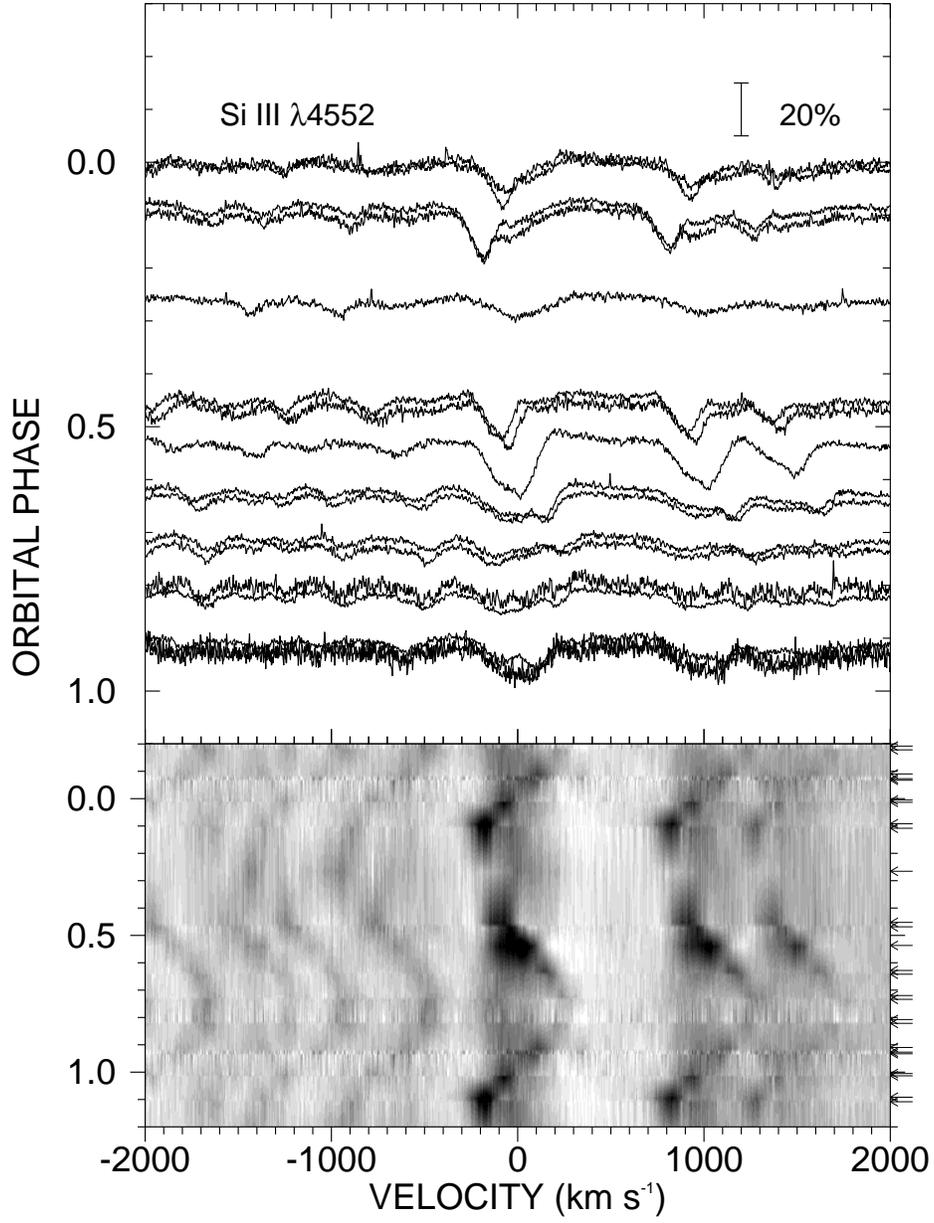}
\caption{The orbital phase variations in the \ion{Si}{3} $\lambda4552$ line
in the same format as Fig.~\ref{SiIV4088}.  All of the triplet \ion{Si}{3} $\lambda4552,4567,4574$
(at 0, 1004, and 1459 km~s$^{-1}$, respectively)
components show dramatic strengthening at the conjunctions and display a broad, 
shallow feature moving in anti-phase as expected for the massive companion.
The absorption feature at $-726$ km~s$^{-1}$ is \ion{He}{2} $\lambda4541$.
\label{SiIII4552}}
\end{figure}

\begin{figure}
\epsscale{.80}
\plottwo{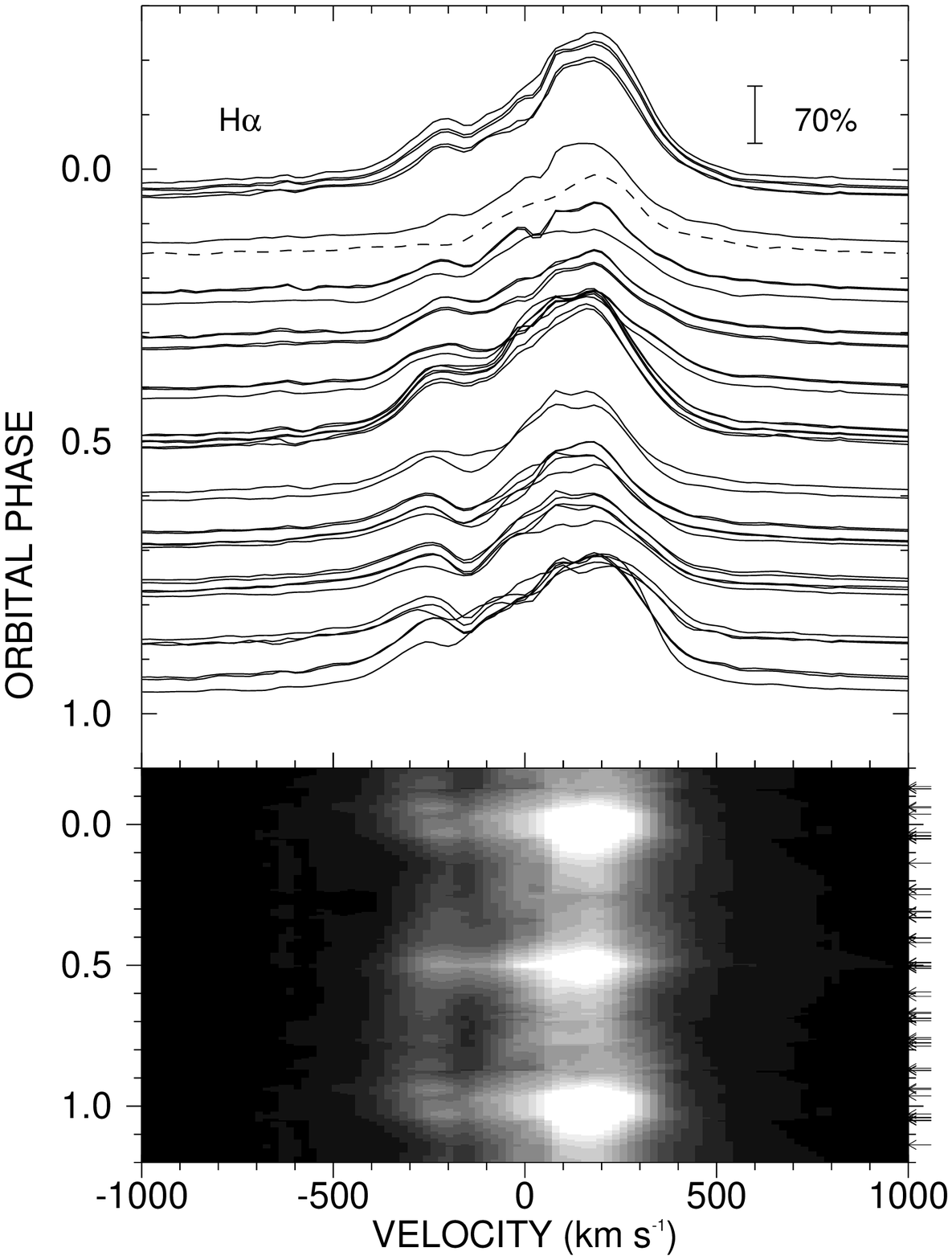}{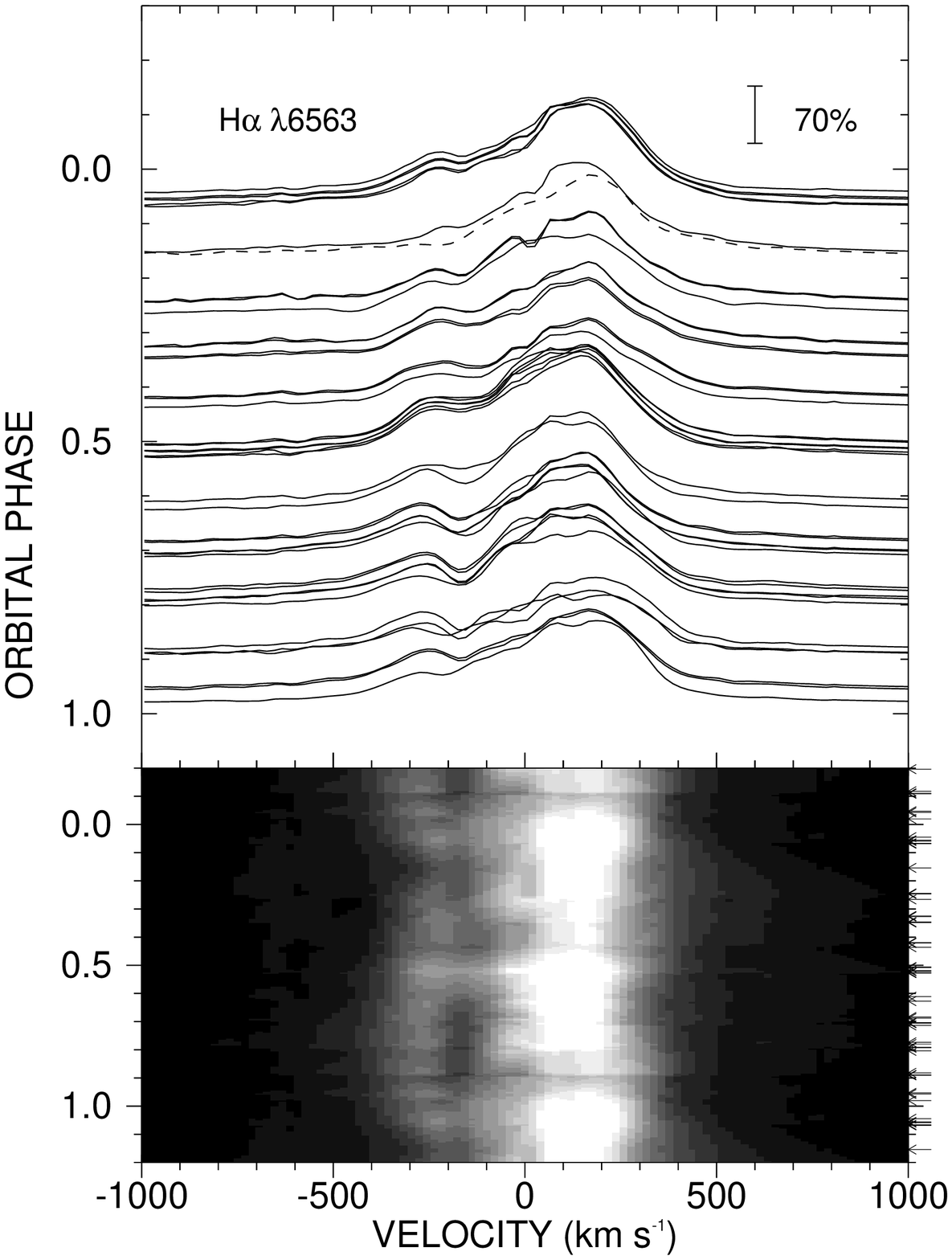}
\caption{The orbital variations in the H$\alpha$ profiles after subtraction of the 
nebular component (based upon the KPNO coud\'{e} feed set of spectra).  
The dashed line at $\phi=0.15$ is a single {\it HST} spectrum of the 
central binary exclusive of the surrounding nebula, and the good match 
to our spectra at that phase indicates that the nebular subtraction method is reliable. 
The left hand panel shows the profiles normalized to the observed continuum (lower near
$\phi = 0.0$ and $\phi = 0.5$) 
while the right hand panel shows the emission rescaled to a constant reference continuum.
The net H$\alpha$ profile is approximately stationary and has high velocity wings that
move in anti-phase to the supergiant's radial velocity curve.
\label{HaCoude}}
\end{figure}

\begin{figure}
\epsscale{.80}
\plotone{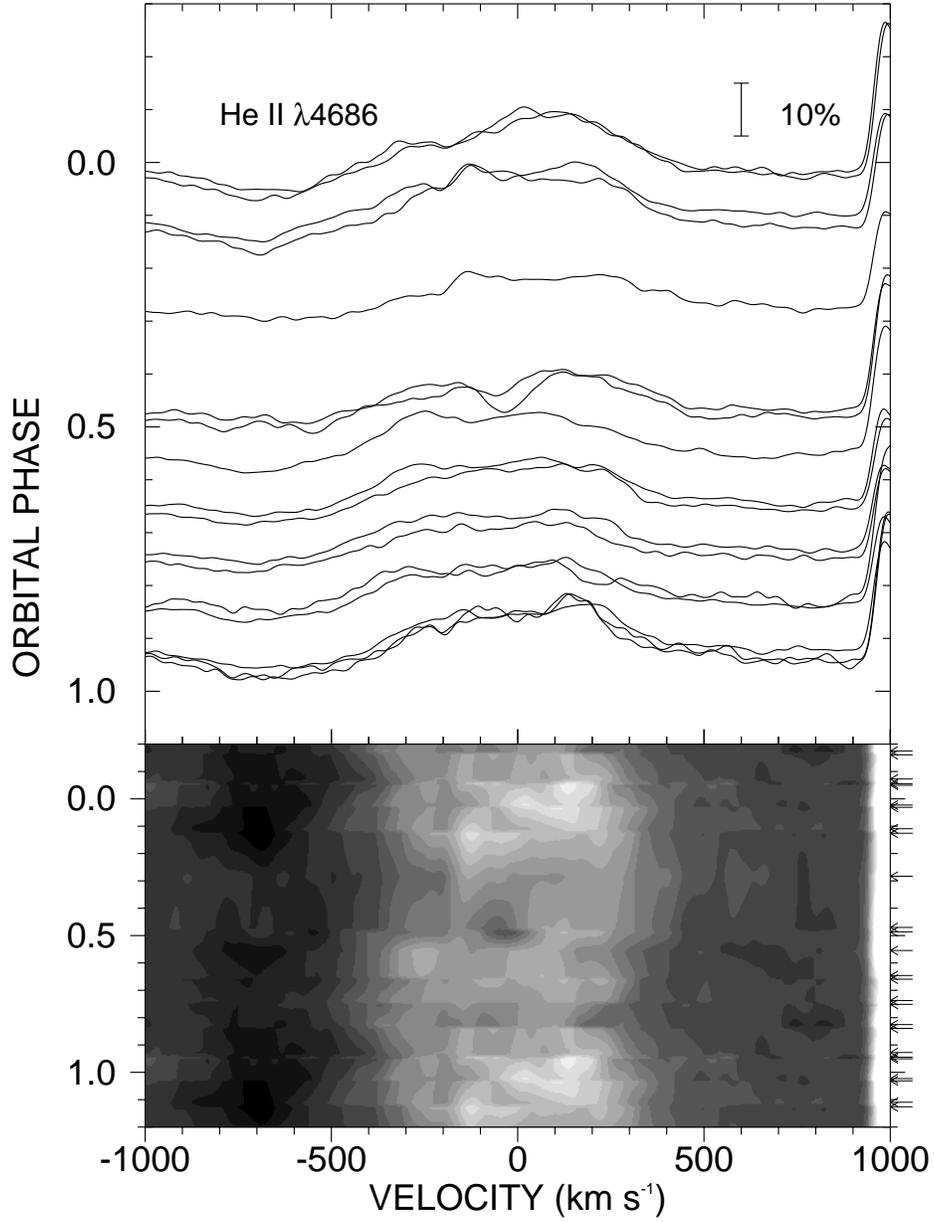}
\caption{The orbital phase variations of the broad and weak \ion{He}{2} $\lambda4686$ 
emission line.  
It appears brighter at $\phi=0.0$ and shows a hint of the anti-phase motion associated 
with the massive companion. 
The nebular emission feature at 1000 km~s$^{-1}$ is [\ion{Fe}{3}] $\lambda4702$.
\label{HeII4686}}
\end{figure}

\begin{figure}
\begin{center}
{\includegraphics[angle=90, height=12cm]{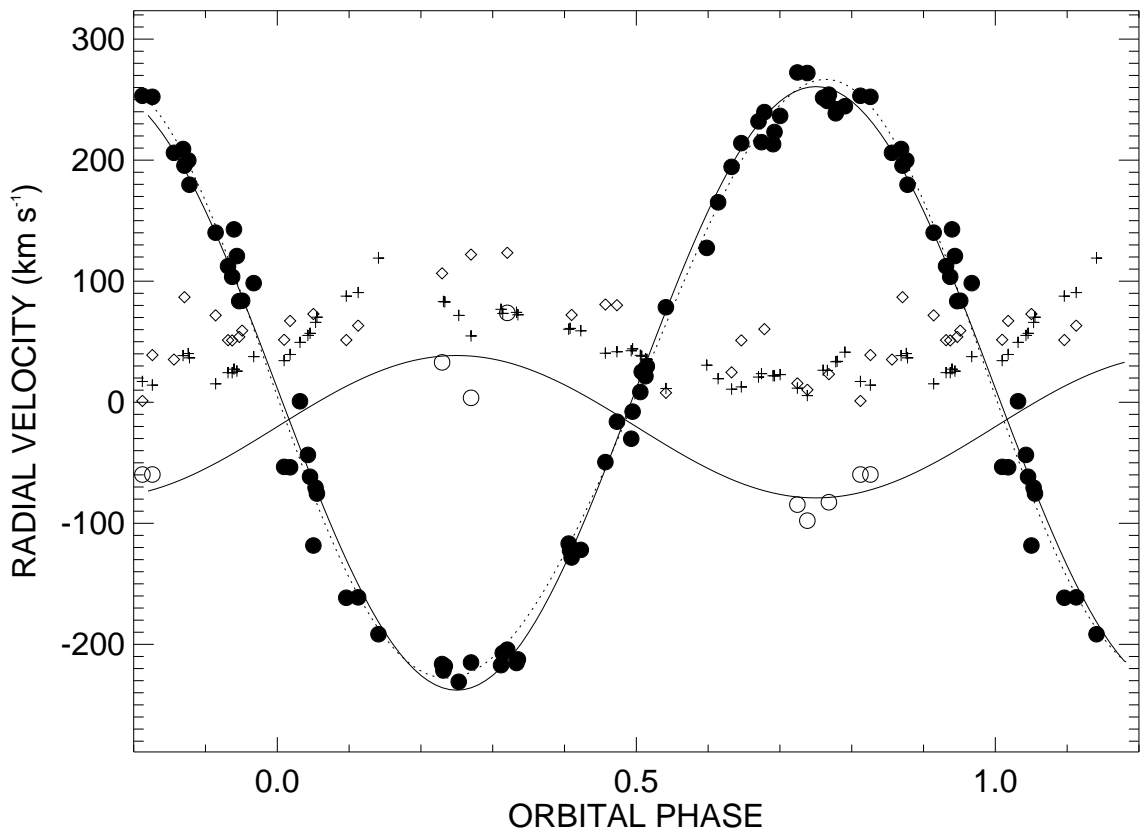}}
\end{center}\caption{The radial velocity curve and measurements for several spectral components
(where $\phi=0.0$ corresponds to supergiant superior conjunction).  
The filled circles represent radial velocities of the supergiant
(with typical errors of 10 km~s$^{-1}$)
while the large amplitude solid line shows the derived velocity curve for the circular solution
and the elliptical solution is shown by the dotted curve (Table~\ref{OrbSol}).
The open circles illustrate the radial velocities for \ion{Si}{3} $\lambda4552$
(with typical errors of 30 km~s$^{-1}$)
and the smaller amplitude solid line shows the constrained fit (Table~\ref{MCRVfits}).
The diamonds represent the radial velocities for the broad \ion{He}{2} $\lambda4686$ emission
(with typical errors of 50 km~s$^{-1}$)
while the plus signs indicate radial velocities for the H$\alpha$ wings
(with typical errors of 2 km~s$^{-1}$).
\label{RV-all}}
\end{figure}

\begin{figure}
\begin{center} 
{\includegraphics[angle=90, height=12cm]{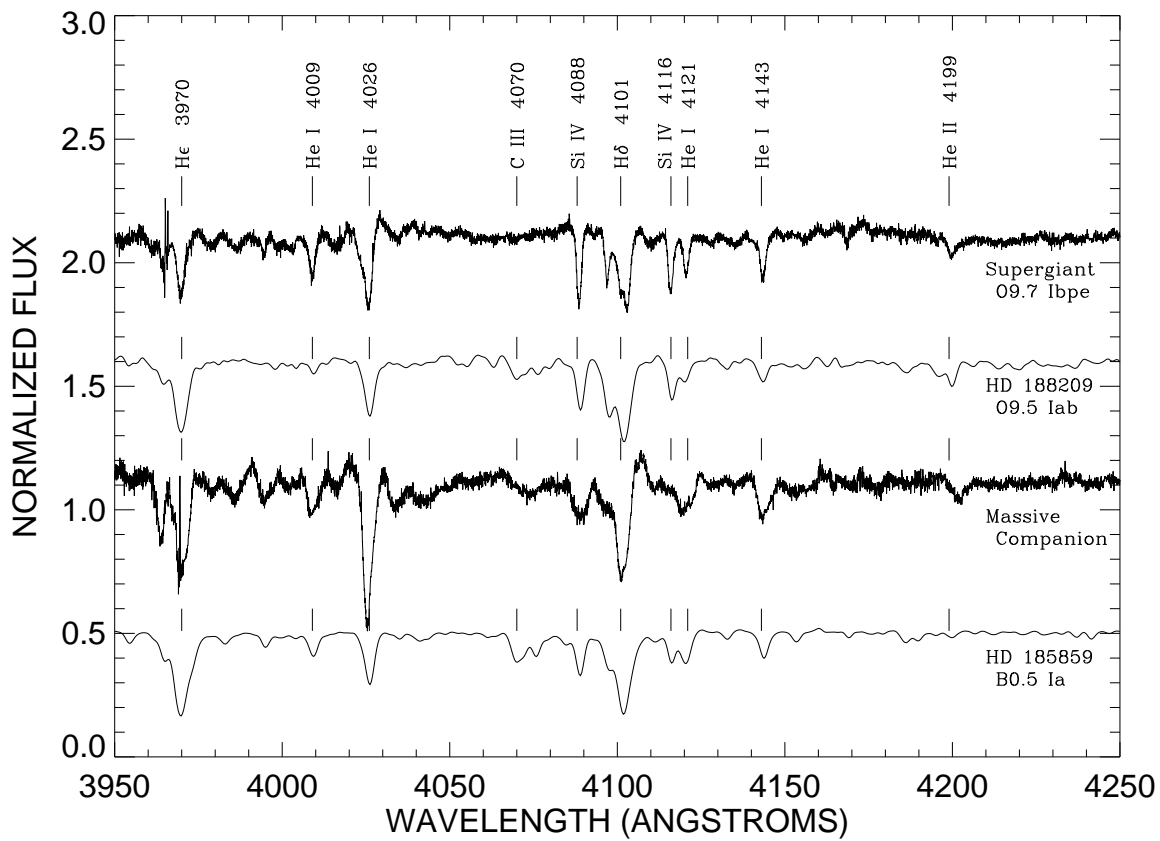}} 
\end{center} 
\caption{Tomographically reconstructed spectra for the supergiant and 
the environment of the massive companion in the spectral range 
$3950 - 4250$ \AA\ plotted using normalized flux.  
These are compared with two single star spectra 
from the atlas of \citet{val04}.   The spectra are offset in flux for clarity. 
The reconstructed spectra display some ``rippling'' near some strong features due 
to residual emission with non-orbital motions.
\label{tomog39}}
\end{figure}

\begin{figure}
\begin{center} 
{\includegraphics[angle=90, height=12cm]{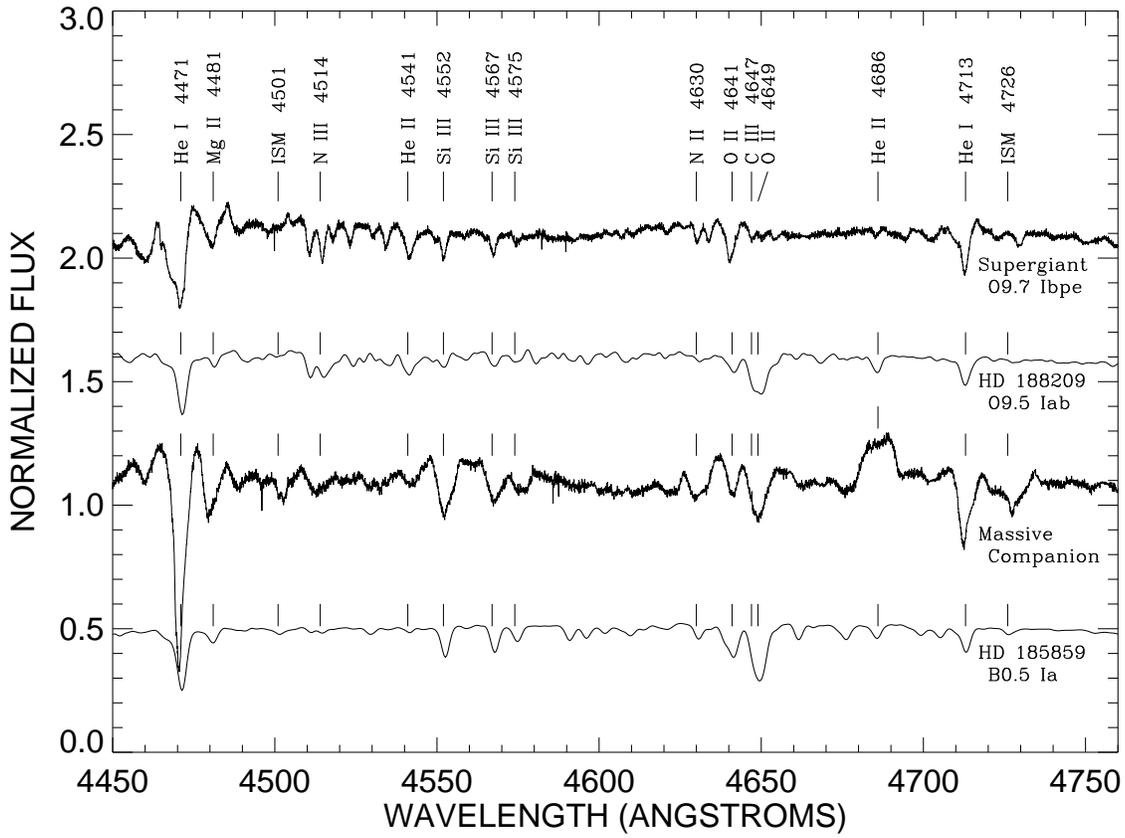}} 
\end{center} 
\caption{Tomographically reconstructed spectra in the range of $4450-4750$ \AA\
(in the same format as Fig.~\ref{tomog39}).  
\label{tomog44}}
\end{figure}

\begin{figure}
\begin{center} 
{\includegraphics[angle=90, height=12cm]{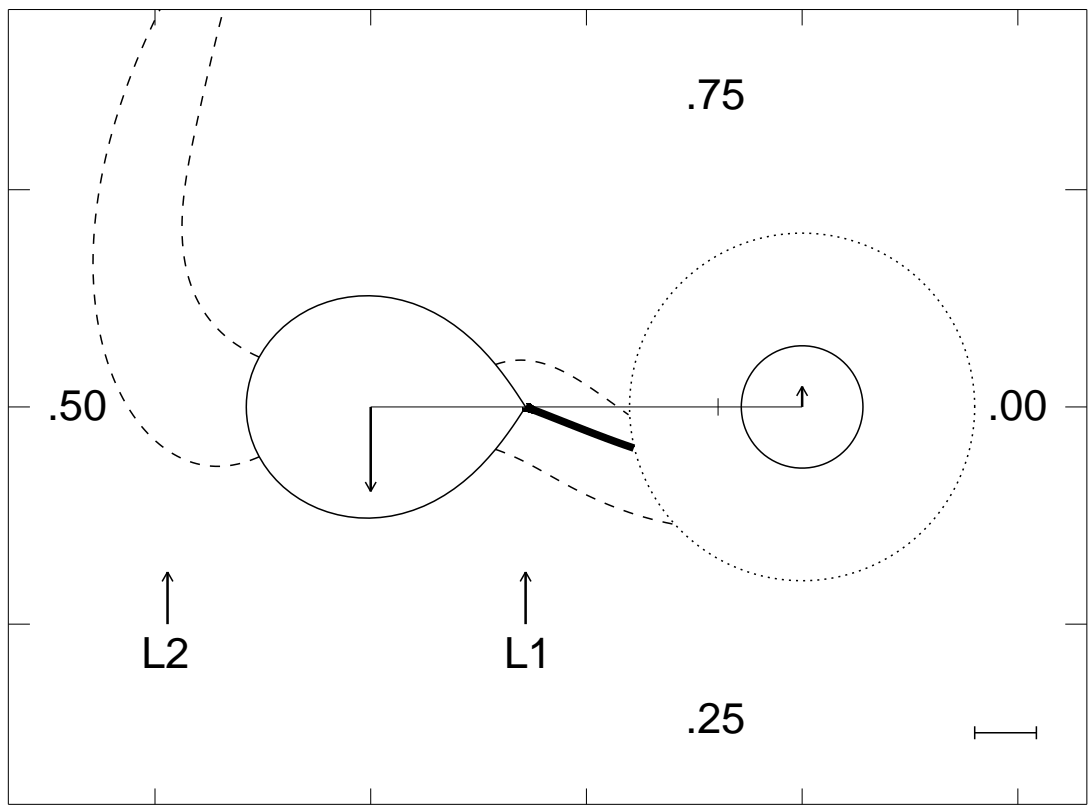}} 
\end{center} 
\caption{Cartoon model of the system as seen from above the orbital plane with the 
observer's orbital phases noted on the periphery and a scale bar denoting 
10$R_\odot$ in the lower right of the diagram.  The Roche-filling 
supergiant appears on the left while the massive companion is shown on the right 
surrounded by a dense accretion disk with the outer boundary shown as the dotted 
line.  The mass loss from the supergiant primarily occurs in two regions: the L1 
region between the stars and the L2 region on the left side of the supergiant.  These 
are delineated by the dashed lines.  The thick solid line represents the classical 
Roche lobe overflow stream from L1 to the disk.  
The tick mark on the axis joining the centers of the two stars marks the center of 
mass, and the arrows at the centers of both stars indicate their orbital velocities. 
The L1 and L2 points lie along the axis joining the two stars at the points indicated.
The L3 point lies off the plot to the right (62$R_\odot$ from the center of the massive
companion).
\label{cartoon}}
\end{figure}

\begin{figure}
\begin{center} 
{\includegraphics[angle=90, height=12cm]{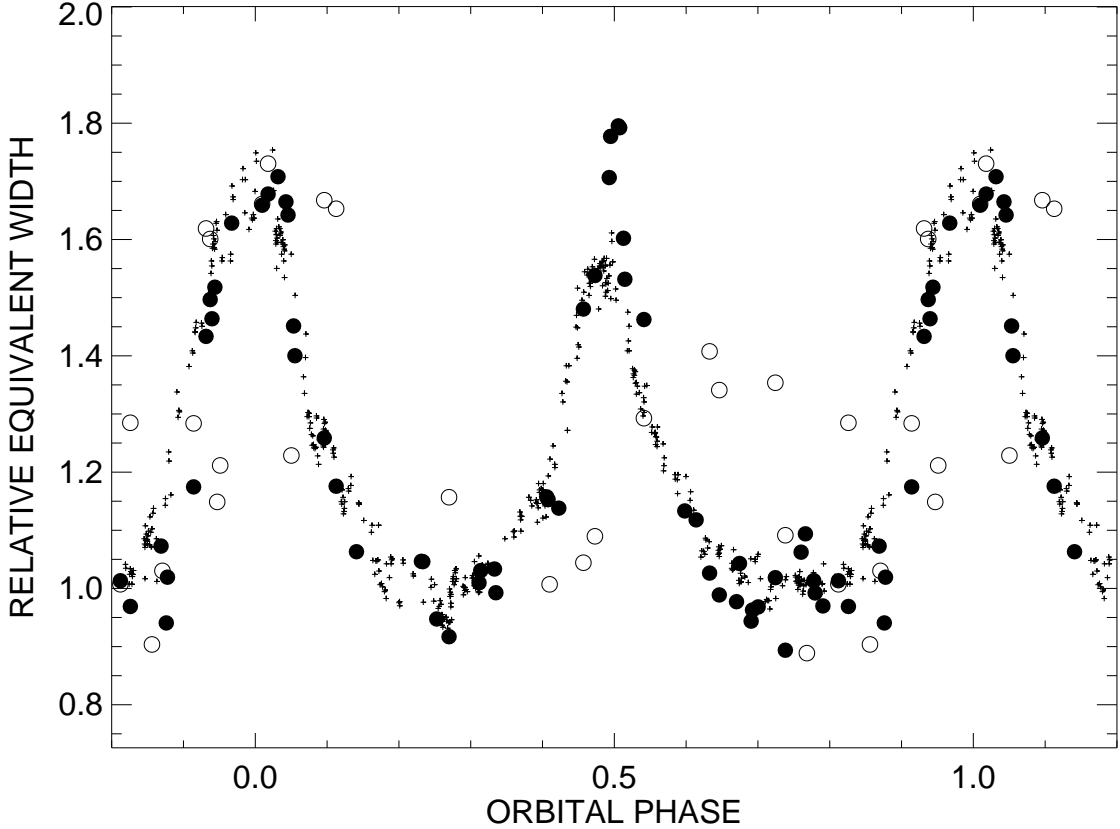}} 
\end{center} 
\caption{A representation of the apparent increases in equivalent width due to the 
changing continuum levels in the eclipsing binary.  The small plus signs mark the 
inverse fluxes in the $V$-band from \citet{djur01}, and these show the predicted trend 
in equivalent width for any uneclipsed emission source.  The filled circles show the 
normalized variation in H$\alpha$ equivalent width that appears to follow the predicted curve 
(indicating that the H$\alpha$ source is not significantly eclipsed).  The open circles show 
the same for \ion{He}{2} $\lambda 4686$.  In this case, the absence of brightening at 
$\phi= 0.5$ suggests that the \ion{He}{2} emission is partially occulted/eclipsed then.
Conservative estimates of the relative equivalent width errors for H$\alpha$ are $4\%$
and those for \ion{He}{2} $\lambda 4686$ are between $4\%$ and $10\%$.
\label{lightcurve}}
\end{figure}

\end{document}